\documentclass[12pt,draftcls,onecolumn]{IEEEtran}

\usepackage{amsmath}
\usepackage{amssymb}
\usepackage{graphicx}
\usepackage{latexsym}
\usepackage{cite}
\usepackage{extraipa}
\usepackage{mathrsfs}
\usepackage{stmaryrd}
\usepackage{overpic}
\usepackage{pdfsync}

\newtheorem{theorem}{Theorem}
\newtheorem{corollary}{Corollary}[theorem]
\newtheorem{lemma}{Lemma}
\newtheorem{proposition}{Proposition}
\newtheorem{definition}{Definition}

\newtheorem{remark}{Remark}
\newtheorem{example}{Example}

\newcommand{\mbf}[1]{\mathbf{#1}}
\newcommand{\set}[1]{\mathscr{#1}}

\newcommand{\markov}{\minuso}
\newcommand{\aux}{U}
\newcommand{\auxb}{W}
\DeclareMathOperator{\expect}{\mathbb{E}} 
\newcommand{\arr}{\set{R}} 
\newcommand{\reals}{\mathbb{R}}

\newcommand{\D}{\mathbf{d}}
\renewcommand{\S}{\set{S}}

\renewcommand{\k}{\mathbf{k}}

\graphicspath{{.}{./Figures/}}

\title{Rate Distortion with Side-Information at Many Decoders}
\author{Roy Timo, Terence Chan and Alex Grant\thanks{
This work was funded by the Australian Research Council Grant DP0880223.}\thanks{R. Timo, T. Chan and A. Grant are with the Institute for Telecommunications Research, University of South Australia. Email: \{roy.timo, terence.chan, alex.grant\}@unisa.edu.au. }}

\begin{document}
\maketitle

\begin{abstract}
We present a new inner bound for the rate region of the $t$-stage successive-refinement problem with side-information. We also present a new upper bound for the rate-distortion function for lossy-source coding with multiple decoders and side-information. Characterising this rate-distortion function is a long-standing open problem, and it is widely believed that the tightest upper bound is provided by Theorem 2 of Heegard and Berger's paper ``Rate Distortion when Side Information may be Absent,'' \emph{IEEE Trans. Inform. Theory}, 1985. We give a counterexample to Heegard and Berger's result.
\end{abstract}


\newpage

\section{Introduction}\label{Sec:1}

\begin{figure}[h]
  \centering
  \includegraphics[width=0.6\columnwidth]{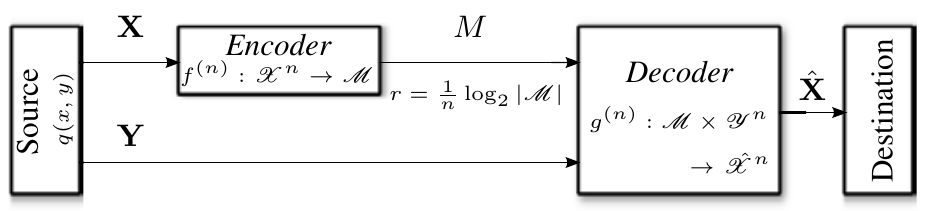}
  \caption{The Wyner-Ziv Problem: $(\mbf{X},$ $\mbf{Y})$ $=$ $(X_1,$ $Y_1)$, $(X_2,$ $Y_2)$, $\ldots$, $(X_n,$ $Y_n)$ is an iid random sequence emitted by a source $q(x,y) = \text{Pr}[X=x,Y=y]$.  The encoder maps $\mbf{X}$ to an index $M$, which belongs to a finite set $\set{M}$, at a rate $r$. Using $M$ and $\mbf{Y}$, the decoder is required to generate a replica $\Hat{\mbf{X}} = \Hat{X}_1,\Hat{X}_2,\ldots,\Hat{X}_n$ of $\mbf{X}$ to within an average distortion $d$, according to~\eqref{Sec:1:Eqn:Distortion-1}. The rate-distortion function $R(d)$ is defined as the smallest rate for which such a reconstruction is possible. A single-letter expression for this function was first given
  in~\cite[Thm. 1]{Wyner-Jan-1976-A}.}
  \label{fig:Wyner-Ziv}
\end{figure}

\PARstart{O}{ne} of the most important and celebrated results in multi-terminal information theory is Wyner and Ziv's solution to the problem of \textit{lossy source coding with side-information at the decoder}~\cite{Wyner-Jan-1976-A} -- the Wyner-Ziv problem (fig.~\ref{fig:Wyner-Ziv}). The main objective of this problem is to find a computable characterisation~\cite[Pg. 259]{Csiszar-1981-B} of the rate-distortion function $R(d)$. This function describes the smallest rate at which the encoder can compress an iid random sequence $\mbf{X}$ so that the decoder, which has side-information $\mbf{Y}$, can produce a replica $\Hat{\mbf{X}}$ of $\mbf{X}$ that satisfies the average distortion constraint
\begin{equation}\label{Sec:1:Eqn:Distortion-1}
\mathbb{E}\left[\frac{1}{n}\sum_{i=1}^n \delta\big(X_i,\Hat{X}_i\big)\right] \leq d\ ,
\end{equation}
where $\delta$ is a real-valued distortion measure~\cite{Cover-1991-B} and $\mathbb{E}[\cdot]$ is the expectation operation. In~\cite[Thm. 1]{Wyner-Jan-1976-A}, Wyner and Ziv famously showed that
\begin{align}
\label{Sec:1:Eqn:Wyner-Ziv-1} R(d) &= \min_U \big\{ I(X;U) - I(U;Y)\big\}\ ,
\end{align}
where the minimization is taken over all choices of an auxiliary random variable $U$ that is jointly distributed with $(X,Y)$ and which satisfies the following two properties: (1) $U$ is conditionally independent of $Y$ given $X$; and (2) there exists a function $\Hat{X}(U,Y)$ with $\mathbb{E}\delta(X,\Hat{X}(U,Y)) \leq d$.

In this paper, we study the following two extensions of the Wyner-Ziv problem: (1) the Wyner-Ziv problem with multiple decoders (fig.~\ref{Sec:1:Fig:mKHB}); and (2) the successive-refinement problem with side-information (fig.~\ref{Sec:1:Fig:tSR}). A brief history of the literature on these problems is as follows.

\subsection{The Wyner-Ziv Problem with $t$-Decoders}

\begin{figure}[h]
\vspace{5mm}
\centering
  \includegraphics[width=.75\columnwidth]{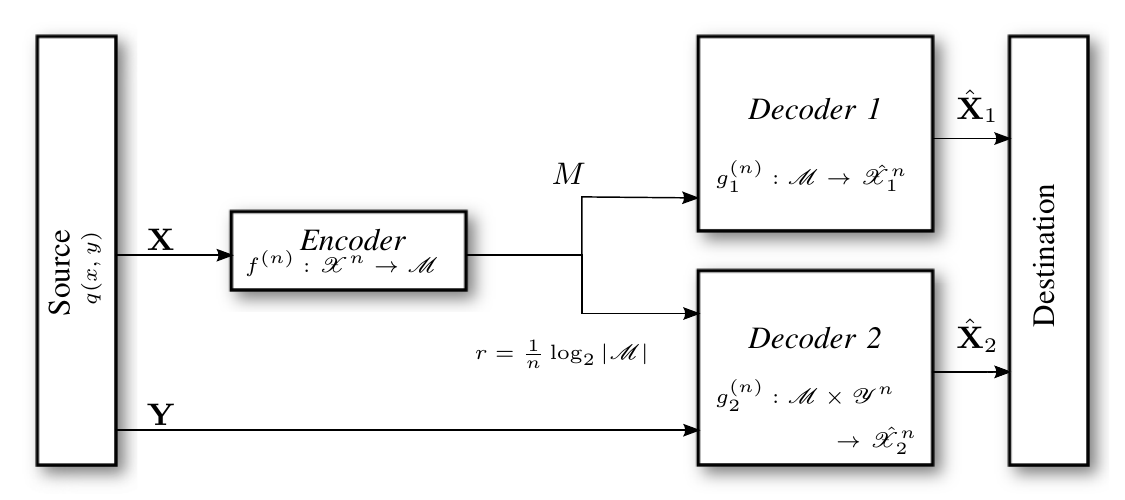}
  \caption{The Kaspi/Heegard-Berger Problem: The encoder compresses $\mbf{X}$ in a manner suitable for two decoders -- one of which has side-information $\mbf{Y}$. The rate-distortion function $R(d)$ defines the smallest rate at which decoders 1 and 2 can generate replicas $\Hat{\mbf{X}}_1 = \Hat{X}_{1,1},\Hat{X}_{1,2},\ldots,\Hat{X}_{1,n}$ and $\Hat{\mbf{X}}_2 = \Hat{X}_{2,1},\Hat{X}_{2,2},\ldots,\Hat{X}_{2,n}$ of $\mbf{X}$ to within average distortions $d_1$ and $d_2$, respectively. A single-letter expression for this function was independently  given in~\cite{Kaspi-Nov-1994-A} and~\cite{Heegard-Nov-1985-A}.}\label{Sec:1:Fig:KHB}
\end{figure}

Suppose that the side-information $\mbf{Y}$ in Figure~\ref{fig:Wyner-Ziv} is unreliable in the sense that it may or may not be available to the decoder. If the encoder does not know \textit{a priori} when $\mbf{Y}$ is available, then Wyner and Ziv's coding argument for~\eqref{Sec:1:Eqn:Wyner-Ziv-1} fails, and a more sophisticated argument is required to exploit $\mbf{Y}$. This observation inspired Kaspi~\cite{Kaspi-Nov-1994-A} in 1980 (published by Wyner on behalf of Kaspi in 1994) as well as Heegard and Berger~\cite{Heegard-Nov-1985-A} in 1985 to independently study the problem shown in fig.~\ref{Sec:1:Fig:KHB} -- the Kaspi/Heegard-Berger problem. As with the Wyner-Ziv problem, the objective of this problem is to characterise the corresponding rate-distortion function $R(d_1,d_2)$. That is, to find the smallest rate such that decoders 1 and 2 can produce replicas $\Hat{\mbf{X}}_1$ and $\Hat{\mbf{X}}_2$ of $\mbf{X}$ to within average distortions $d_1$ and $d_2$, respectively. To this end, Heegard and Berger~\cite[Thm. 1]{Heegard-Nov-1985-A} showed that\footnote{Kaspi's result, \cite[Thm. 2]{Kaspi-Nov-1994-A}, gives an alternative characterisation of $R(d_1,d_2)$ that uses one auxiliary random variable.}
\begin{equation*}
R(d_1,d_2) = \min_{U,W} \big\{ I\left(X;\auxb \right) +
I\left(X;\aux \mid Y, \auxb \right) \big\}\ ,
\end{equation*}
where the minimization is taken over all choices of two auxiliary random variables, $\aux$ and $\auxb$, that are jointly distributed with $(X,Y)$ and which satisfy the following two properties: (1)
$(\aux,\auxb)$ is conditionally independent of $Y$ given $X$; and (2)
there exist functions $\Hat{X}_{1}(\auxb)$ and $\Hat{X}_{2}(Y,\aux,\auxb)$ with $\mathbb{E}\delta(X,\Hat{X}_1(\auxb)) \leq d_1$ and $\mathbb{E}\delta(X,\Hat{X}_2(Y,\aux,\auxb)) \leq d_2$, respectively.

\begin{figure}
\vspace{5mm}
  \centering
\includegraphics[width=.75\columnwidth]{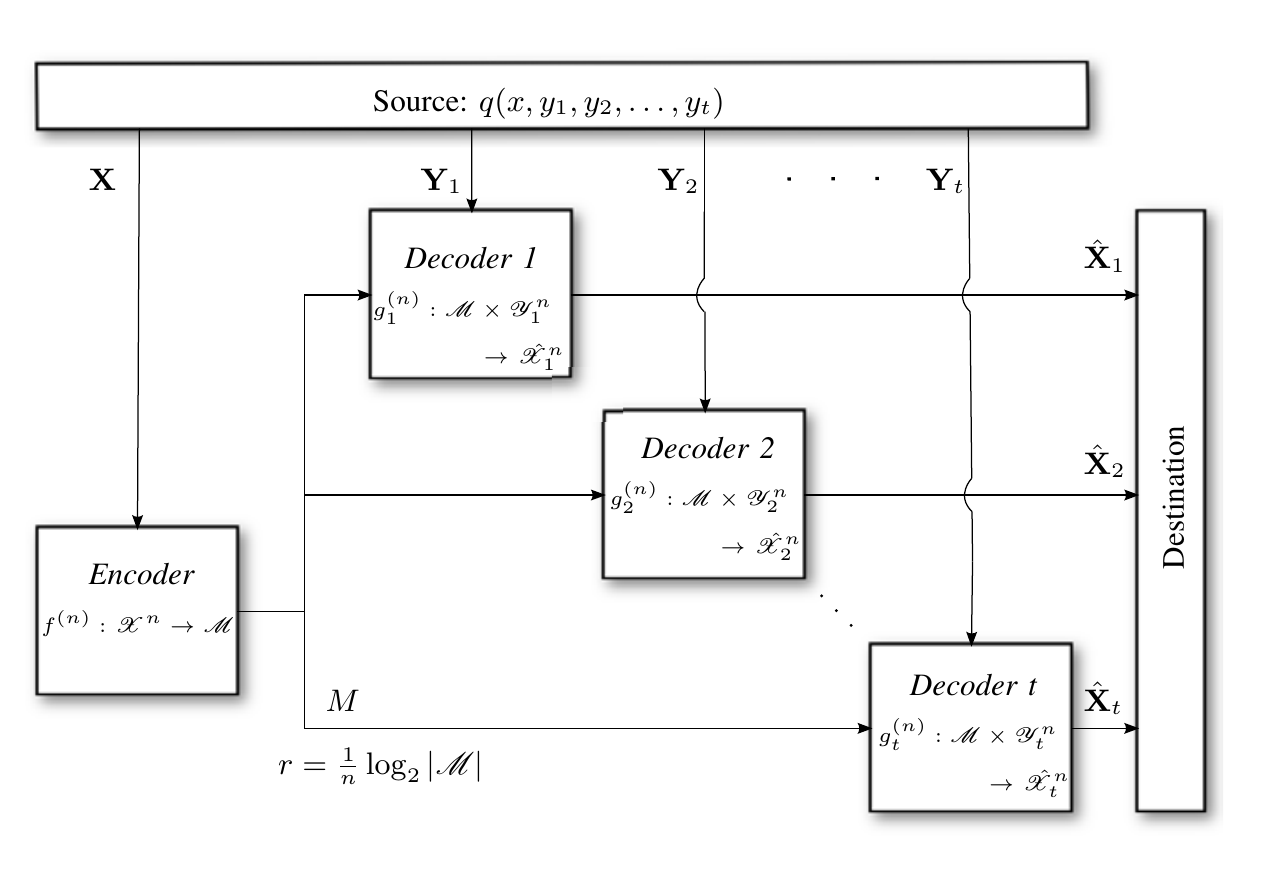}
  \caption{The Wyner-Ziv problem with $t$-decoders. The encoder compresses $\mbf{X}$ in a manner suitable for $t$-decoders -- each of which has different side-information. The rate-distortion function $R(\mbf{d})$, where $\mbf{d} = (d_1,d_2,\dots,d_t)$, defines the smallest rate at which decoder $l$, for all $l =1,2,\ldots,t$, can generate a replica $\Hat{\mbf{X}}_l$ of $\mbf{X}$ to within an average distortion $d_l$. This problem is open for $t \geq 2$. We present an upper bound for $R(\mbf{d})$ in Theorem~\ref{Sec:4:Thm:HB}.}\label{Sec:1:Fig:mKHB}
\end{figure}

The Kaspi/Heegard-Berger problem in Figure~\ref{Sec:1:Fig:KHB} was further generalised by Heegard and Berger in \cite[Sec. VII]{Heegard-Nov-1985-A} to the problem shown in
Figure~\ref{Sec:1:Fig:mKHB}. There are $t$-decoders, each with different side-information, and the objective is to characterise the corresponding rate-distortion function $R(\mbf{d})$. Unfortunately, this function has eluded characterisation for all but a few special cases. For example, Heegard and Berger~\cite[Thm. 3]{Heegard-Nov-1985-A} have characterised $R(\mbf{d})$ for stochastically degraded side-information\footnote{The joint probability distribution of $(X,$ $Y_1,$ $Y_2,$ $\ldots,$ $Y_t)$ can be manipulated to form the Markov chain $X \minuso Y_t \minuso Y_{t-1} \minuso \cdots \minuso Y_1$ without altering $R(\mbf{d})$. We discuss this problem in detail in Section~\ref{Sec:2:C}.}; Tian and Diggavi~\cite{Tian-Aug-2007-A,Tian-Dec-2008-A} have characterised $R(\mbf{d})$ for a quadratic Gaussian source with jointly Gaussian side-information; and Sgarro's result~\cite[Thm. 1]{Sgarro-Mar-1977-A} subsumes the corresponding lossless problem. Notwithstanding this difficulty, however, this problem has helped stimulate a number of important results~\cite{Kaspi-Nov-1994-A,Steinberg-Aug-2004-A,Tian-Oct-2006-C,Tian-Dec-2008-A,Tian-Aug-2007-A}.

In \cite[Thm. 2]{Heegard-Nov-1985-A}, Heegard and Berger claimed that a certain functional, $R_0(\mbf{d})$, is an upper bound for $R(\mbf{d})$. (The expression for $R_0(\mbf{d})$ is given in equation~\eqref{Sec:2:Eqn:HB-Upper-Bound} of Section~\ref{Sec:2}; however, this expression requires notation from Section~\ref{Sec:2}.) For twenty-five years, $R_0(\mbf{d})$ has been universally considered to be the tightest upper bound for $R(\mbf{d})$ in the literature. In Example~\ref{Sec:2:Counterexample} of Section~\ref{Sec:2}, we present a counterexample to~\cite[Thm. 2]{Heegard-Nov-1985-A} that shows $R_0(\mbf{d})$ is \textit{not} an upper bound for $R(\mbf{d})$. The invalidity of~\cite[Thm. 2]{Heegard-Nov-1985-A} is by no means obvious as it involves a difficult minimization over $(2^t - 1)$-auxiliary random variables. Indeed, we note that this theorem has been cited with modest frequency in the literature, and all the while this error appears to have gone unnoticed. We present a new upper bound for  $R(\mbf{d})$ in Theorem~\ref{Sec:4:Thm:HB} of Section~\ref{Sec:4}.

\subsection{The Successive-Refinement Problem with Side-Information}

\begin{figure}[h]
\vspace{5mm}
  \centerline{\includegraphics[width=.75\columnwidth]{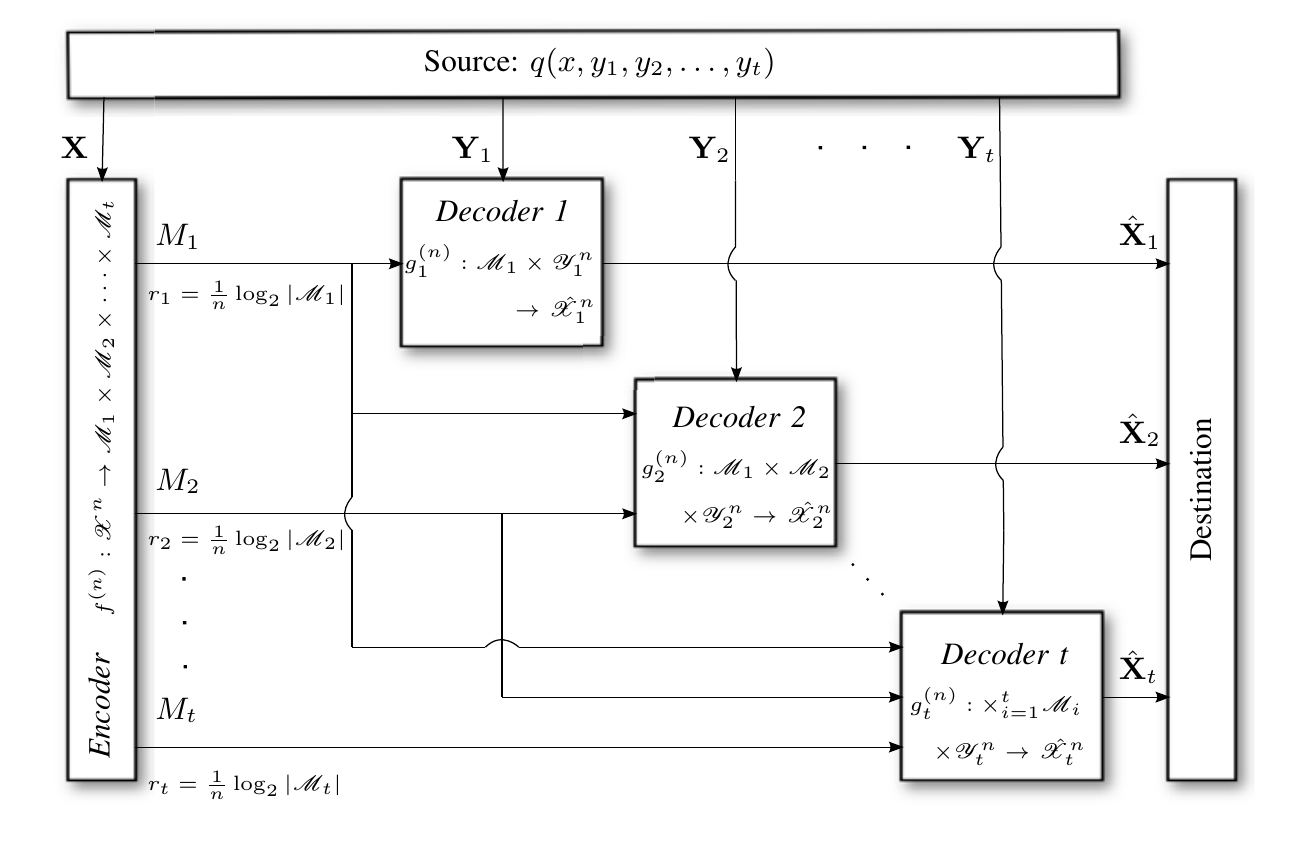}\\}
  \caption{The successive-refinement problem with $t$ stages. The encoder compresses $\mbf{X}$ in $t$-stages. At stage $l$, decoder $l$ generates a replica $\Hat{\mbf{X}}_l$ of $\mbf{X}$. This problem is open for $t \geq 2$. We present an inner bound for the admissible-rate region in Section~\ref{Sec:3} (Theorem~\ref{Sec:3:Thm:Successive-Refinement}). }\label{Sec:1:Fig:tSR}
\end{figure}

The aforementioned counterexample led us to study the $t$-stage (or, $t$-decoder) successive-refinement problem shown in Figure~\ref{Sec:1:Fig:tSR}. The encoder maps $\mbf{X}$ to $t$ indices: $M_1, M_2, \dots, M_t$. It is required that decoder $l$ uses indices $M_1$ through $M_l$ together with its side-information $\mbf{Y}_l$ to produce a replica $\Hat{\mbf{X}}_l = X_{l,1},X_{l,2},\ldots,X_{l,n}$ of $\mbf{X}$ to within an average distortion $d_l$. The objective of this problem is to characterise the resulting admissible-rate region $\set{R}(\mbf{d})$. That is, to determine the set of all rate tuples $\mbf{r} = (r_1,r_2,\ldots,r_t)$ for which each decoder can reconstruct $\mbf{X}$ to within its desired distortion level.

Assuming the side-information is stochastically degraded, Steinberg and Merhav~\cite{Steinberg-Aug-2004-A} characterised $\set{R}(d_1,d_2)$ for $t=2$ decoders. Shortly thereafter, Tian and Diggavi~\cite{Tian-Aug-2007-A} extended this problem to $t$-decoders and proved the following result.

\medskip

\begin{proposition}\label{Sec:1:Pro:Degraded-Side-Information}
If the side-information is stochastically degraded, then $\set{R}(\mbf{d})$ is equal to the set of all rate tuples $\mbf{r}$ for which there exists $t$ auxiliary random variables $U_1$, $U_2$, $\ldots$, $U_t$ such that
\begin{equation*}
\sum_{k=1}^l r_k \geq \sum_{k=1}^l I\big(X;U_k\big|U_1,U_2,\ldots,U_{k-1},Y_k\big)\ ,
\end{equation*}
for all $l = 1,2,\ldots,t$, where
\begin{enumerate}
\item $(U_1,U_2,\ldots,U_t)$ is conditionally independent of $(Y_1,$ $Y_2,$ $\ldots,$ $Y_t)$ given $X$; and
\item there exist $t$ functions $\Hat{X}_l(U_l,Y_l)$, $l = 1,2,\ldots,t$, with $\mathbb{E}\delta_l(X,\Hat{X}(U_l,Y_l)) \leq d_l$.
\end{enumerate}
\end{proposition}

\medskip

More recently, Tian and Diggavi~\cite{Tian-Dec-2008-A} gave the following non-trivial inner bound for $\set{R}(d_1,d_2)$ under the assumption that $X$ and $Y_2$ are conditionally independent given $Y_1$ -- the scalable side-information source coding problem.  Note, this conditional independence is the reverse of the stochastic degradedness used in Proposition~\ref{Sec:1:Pro:Degraded-Side-Information}.

\medskip

\begin{proposition}\label{Sec:1:Pro:SI-Scalable}
If $X$ and $Y_2$ are conditionally independent given $Y_1$, then a rate pair $(r_1,r_2)$ is $(d_1,d_2)$-admissible if there exists three auxiliary random variables, $U_{12}$, $U_1$ and $U_2$, such that
\begin{align*}
r_1 &\geq I\big(X;U_1,U_{12}\big|Y_1\big)\\
r_1 + r_2 &\geq I\big(X;U_2,U_{12}\big|Y_2) + I\big(X;U_1\big|Y_1,U_{12}\big)\ ,
\end{align*}
where
\begin{enumerate}
\item $(U_{12},U_1,U_2)$ is conditionally independent of $(Y_1,Y_2)$ given $X$;
\item there exist functions $\Hat{X}_1(U_1,Y_1)$ and $\Hat{X}_2(U_2,Y_2)$ such that $\mathbb{E}\delta_1(X,\Hat{X}_1(U_1,Y_1))\leq d_1$ and $\mathbb{E}\delta_2(X,$ $\Hat{X}_2(U_2,Y_2))$ $\leq$ $d_2$.
\end{enumerate}
\end{proposition}

\medskip

We present a new inner bound for $\set{R}(\mbf{d})$ for the general $t$-decoder problem with arbitrarily correlated side-information in Theorem~\ref{Sec:3:Thm:Successive-Refinement} of Section~\ref{Sec:3}.

\subsection{Paper Outline $\&$ Notation}

In Section~\ref{Sec:2}, we formally define $\set{R}(\mbf{d})$ and $R(\mbf{d})$ and give the counterexample to~\cite[Thm. 2]{Heegard-Nov-1985-A}. In Sections~\ref{Sec:3} and~\ref{Sec:4}, we respectively present new achievability results for $\set{R}(\mbf{d})$ and $R(\mbf{d})$. We describe a new lossless source coding problem in Section~\ref{Sec:5}, and the paper is concluded in Section~\ref{Sec:6}.

The non-negative real numbers and the natural numbers are written as $\mathbb{R}_+$ and $\mathbb{N}$, respectively. For $s,t \in \mathbb{N}$ with $s\leq t$, we let $[s,t] \triangleq \{s,s+1,s+2,\dots,t\}$. When $s = 1$, we drop $s$, i.e. $[t] \triangleq \{1,2,\ldots,t\}$. Proper subsets and subsets are identified by $\subset$ and $\subseteq$, respectively. Random variables and random sequences are identified by upper case and bolded uppercase letters, respectively. For example, $\mbf{X} = X_1,X_2,\ldots,X_n$ denotes the random sequence to be replicated at the decoders, and $\mbf{Y}_l = Y_{l,1},Y_{l,2},\ldots,Y_{l,n}$ denotes the side-information at decoder $l$. The letter $U$ is always used to represent auxiliary random variables. The alphabets of random variables are identified by matching calligraphic typeface, e.g. $\set{X}$ and $\set{U}$ are the respective alphabets of $X$ and $U$. A generic element of an alphabet is identified by a matching lowercase letter, e.g. $x \in \set{X}$ and $u \in \set{U}$. The Cartesian product operation is denoted by $\times$, e.g. $\set{X} \times \set{Y}$. The $t$-fold Cartesian product of a single alphabet/set is identified with a superscript, e.g. $\set{X}^t$ and $\mathbb{R}_+^t$. Tuples from product spaces are identified by boldfaced lowercase letters, e.g. $\mbf{x} = (x_1,x_2,\ldots,x_n) \in \set{X}^n$.

For notational convenience, the same letter is used to represent a joint pmf and its marginals, e.g. if $(X,Y)$ on $\set{X} \times \set{Y}$ is defined by $p(x,y) \triangleq \text{Pr}[X=x,Y=y]$, then $p(x) \triangleq \sum_{x\in\set{X}} p(x,y)$. The symbol $\minuso$ is used to denote Markov Chains, e.g. if $(X,Y,Z)$ on $\set{X} \times \set{Y} \times \set{Z}$ is defined by $p(x,y,z) \triangleq \text{Pr}[X=x,Y=y,Z=z]$ where
\begin{equation*}
p(x,y,z) = \left\{
             \begin{array}{ll}
               p(x,y)p(y,z)/p(y), & \hbox{ if } p(y) > 0 \\
               0, & \hbox{ otherwise,}
             \end{array}
           \right.
\end{equation*}
then we write $X \minuso Y \minuso Z\ [p]$. Mutual information and entropy are written in the standard fashion~\cite{Cover-1991-B} using $I$ and $H$, respectively. We sometimes use subscripts for $I$ and $H$ to emphasize that random variables under consideration are defined by a particular pmf, e.g. if $(X,Y)$ is defined by $p(x,y) = \text{Pr}[X=x,Y=y]$, then we write $I_p(X;Y)$.

\section{Definitions $\&$ Counterexample}\label{Sec:2}

\subsection{Successive Refinement with Side-Information}\label{Sec:2:A}
Consider Figure~\ref{Sec:1:Fig:tSR}. Let $\set{X}$, $\set{Y}_1$, $\set{Y}_2$, $\ldots$, $\set{Y}_t$ be finite alphabets and set $\set{Y}^* \triangleq \set{Y}_1 \times \set{Y}_2 \times \cdots \times \set{Y}_t$. Let
\begin{equation*}
\big(\mbf{X},\mbf{Y}_{1},\mbf{Y}_{2},\dots,\mbf{Y}_{t}\big)
\triangleq \Big\{\big(X_i, Y_{1,i},Y_{2,i},\dots,Y_{t,i})\Big\}_{i=1}^n
\end{equation*}
denote $n$ $(t+1)$-tuples of random variables that are drawn in an iid manner from $\set{X}\times \set{Y}^*$ according to a generic pmf $q$, where
\begin{equation*}
q\left(x,y_{1},\dots,y_{t}\right) \triangleq \Pr\big[X_1 = x_1, Y_{1} = y_{1},\ldots, Y_{t} = y_{t}\big].
\end{equation*}

We assume that $\mbf{X} = X_1,X_2,\ldots,X_n$ is known to encoder and $\mbf{Y}_l = Y_{l,1},Y_{l,2},\ldots,Y_{l,n}$ is known to decoder $l$. The encoder compresses $\mbf{X}$ with
\begin{equation*}
f^{(n)} : \set{X}^n \rightarrow \set{M}_1 \times \set{M}_2 \times \cdots \times \set{M}_t\ ,
\end{equation*}
where $\set{M}_1$, $\set{M}_2$, $\ldots$, $\set{M}_t$ are finite sets. The resulting $t$ indices
\begin{equation*}
(M_{1},M_{2},\ldots,M_{t}) = f^{(n)}\left(\mbf{X}\right)
\end{equation*}
are sent over channels $1$ through $t$, respectively. The rate of the encoder on channel $l$  (in bits per source symbol) is given by
\begin{equation*}
\kappa_l^{(n)} \triangleq \frac{1}{n} \log_2 |\set{M}_l|\ ,
\end{equation*}
where $|\set{M}_l|$ is the cardinality of $\set{M}_l$.

Consider decoder $l$. Let $\Hat{\set{X}}_{l}$ be a finite reconstruction alphabet, and let
\begin{equation*}
\delta_{l} : \set{X} \times \Hat{\set{X}}_{l} \rightarrow \reals_+
\end{equation*}
be a per-letter distortion measure. Observe that $\Hat{\set{X}}_l$ and $\delta_l$ can be different to those used at the other decoders. We assume that $\delta_l$ is normal\footnote{It is possible to remove this assumption and extend the results of this paper to general reconstruction alphabets and per-letter distortion measures using the procedure given in~\cite[Sec. 9.1]{Yeung-2002-B}.} in sense that $\delta_l(x,\Hat{x}^*(x)) = 0$ for all $x \in \set{X}$, where
\begin{equation*}
\Hat{x}^*(x) \triangleq \underset{\Hat{x} \in \Hat{\set{X}}}{\operatorname{argmin}}\  \delta_l(x,\Hat{x})\ .
\end{equation*}
This decoder is required to generate a replica $\Hat{\mbf{X}}_{l}\triangleq \Hat{X}_{l,1},\Hat{X}_{l,2},\ldots,\Hat{X}_{l,n}$ of  $\mbf{X}$ using
\begin{equation*}
g^{(n)}_l : \set{M}_1 \times \set{M}_2 \cdots \times \set{M}_l \times \set{Y}^n_{l} \rightarrow \Hat{\set{X}}_{l}^n\ ;
\end{equation*}
that is,
\begin{equation*}
\Hat{\mbf{X}}_{l} = g_l^{(n)}(M_1, M_2, \ldots,M_l,\mbf{Y}_{l})\ .
\end{equation*}
Finally, the quality of this replica is measured by the average distortion
\begin{equation*}
\Delta_l^{(n)} \triangleq  \expect\left[ \frac{1}{n}\sum_{i=1}^n  \delta_l(X_i,\Hat{X}_{l,i})\right]\ .
\end{equation*}

\medskip

\begin{definition}[$\D$-Admissible Rates]\label{Sec:2:Def:d-admissible}
Suppose $\D$ $=$ $(d_1,$ $d_2,$ $\ldots,$ $d_t)$ $\in$ $\reals_+^t$. A rate tuple $\mbf{r}$ $=$ $(r_1$, $r_2$, $\ldots$, $r_t)$ $\in$ $\reals_+^t$ is said to be {\em $\D$-admissible} if, for arbitrary $\epsilon > 0$, there exists an $n_\epsilon \in \mathbb{N}$, an encoder $f^{(n_\epsilon)}$ and $t$-decoders $g^{(n_\epsilon)}_1$, $g^{(n_\epsilon)}_2$, $\ldots$, $g^{(n_\epsilon)}_t$ such that
\begin{align*}
d_l + \epsilon &\geq \Delta_l^{(n_\epsilon)},\quad  \forall l \in [t], \text{ and}\\
r_l + \epsilon &\geq \kappa_l^{(n_\epsilon)}, \quad \forall l \in [t] .
\end{align*}
We let $\arr(\D)$ denote the set of all $\D$-admissible rate tuples.
\end{definition}

\medskip

We note that Definition~\ref{Sec:2:Def:d-admissible} matches Tian and Diggavi~\cite{Tian-Aug-2007-A} in that the $l^{\text{th}}$ channel (or, refinement) rate $\kappa_l^{(n)}$ is characterised in an individual (or, incremental) manner. In contrast, Steinberg and Merhav~\cite{Steinberg-Aug-2004-A} define the $l^{\text{th}}$ refinement rate in a cumulative manner, e.g. $(1/n)\log(|\set{M}_1||\set{M}_2|\cdots|\set{M}_l|)$. We also note that $\arr(\D)$ is dependent on the successive-refinement decoding order~\cite{Tian-Dec-2008-A}. That is, if we interchange decoders (keeping the same side-information and distortion constraints at each decoder), then $\set{R}(\mbf{d})$ will change.

We conclude this section with a summary of some fundamental properties of $\set{R}(\mbf{d})$. These properties can all be deduced directly from Definition~\ref{Sec:2:Def:d-admissible}. See ~\cite{Gray-Nov-1974-A,Effros-Sep-1999-A,Steinberg-Aug-2004-A,Tian-Aug-2007-A,Tian-Jul-2009-C} for similar discussions.

\medskip

\begin{proposition}\label{Sec:2:Pro:Marginal-Property}
The region $\set{R}(\mbf{d})$ is completely defined by the pair-wise marginal distributions of $X$ with each side-information. Let $q'$ and $q''$ be pmfs on $\set{X}\times\set{Y}^*$, and let $\set{R}(\mathbf{d})[q']$ and $\set{R}(\mathbf{d})[q'']$ denote their respective $\mbf{d}$-admissible rate regions (assuming the same distortion measures). If $q'(x,y_l) = q''(x,y_l)$ for all $(x,y_l) \in \set{X} \times \set{Y}_l$ and $l \in [t]$, then $\set{R}(\mathbf{d})[q'] = \set{R}(\mathbf{d})[q'']$.
\end{proposition}

\medskip

\begin{proposition}\label{Sec:2:Pro:Convex}
The region $\set{R}(\mbf{d})$, for every $\mbf{d} \in \reals_+^t$, is a closed convex subset of $\reals_+^t$ that is uniquely determined by its lower boundary
\begin{equation*}
\Big\{\mbf{r} \in \set{R}(\mbf{d})\ :\ \forall \tilde{\mbf{r}} \in \set{R}(\mbf{d}),\ \tilde{r}_l \leq r_l\ , \forall l \in [t]\ \Rightarrow \tilde{r}_l = r_l\ \forall l \in [t]\Big\}\ .
\end{equation*}
\end{proposition}

\medskip

\begin{proposition}\label{Sec:2:Pro:Sum-Incremental}
The region $\set{R}(\mbf{d})$ is sum incremental in the sense that rate can always be transferred from higher-index channels to lower-index channels. If $\mbf{r} \in \set{R}(\mbf{d})$, then
\begin{equation}\label{Sec:2:Eqn:Latent}
\set{R}(\mbf{d}) \supseteq \set{L}(\mbf{r}) \triangleq \left\{\tilde{\mbf{r}} \in \reals_+^t\ :\ \sum_{k=1}^l \tilde{r}_k \geq \sum_{k=1}^l r_k\ \quad \forall l \in [t]\right\} \ .
\end{equation}
\end{proposition}

\bigskip

We note in passing that Proposition~\ref{Sec:2:Pro:Sum-Incremental} also holds in a more universal setting. Suppose $\mbf{r} \in \reals_+^t$. Consider all combinations of the source distribution, distortion measures and distortion tuple (e.g., $\tilde{\set{X}}$, $\tilde{\set{Y}}^*$, $\tilde{q}$, $\{\tilde{\delta}_l\}_{l=1}^t$ and $\tilde{\mbf{d}} \in \reals_+^t$) such that the resulting $\tilde{\mbf{d}}$-admissible rate region $\set{R}(\tilde{\mbf{d}})[\tilde{q}]$ contains $\mbf{r}$. The proposition shows that $\set{L}(\mbf{r})$ is an inner bound for every such region. In addition, it can be shown that $\set{L}(\mbf{r})$ is maximal in the sense that $\set{L}(\mbf{r}) = \set{R}(\tilde{\mbf{d}})[\tilde{q}]$ for some choice of $\tilde{\set{X}}$, $\tilde{\set{Y}}^*$, $\tilde{q}$, $\{\tilde{\delta_l}\}$ and $\tilde{\mbf{d}}$. Therefore, the $\mbf{d}$-admissibility of $\tilde{\mbf{r}} \notin \set{L}(\mbf{r})$ cannot be inferred from the $\mbf{d}$-admissibility of $\mbf{r}$ without specific consideration of the source distribution, distortion measures and distortion tuple. For this reason, $\set{L}(\mbf{r})$ can be called the latent admissible rate region implied by $\mbf{r}$. See, for example, \cite{Tian-Jul-2009-C}.

We give an inner bound for $\set{R}(\mbf{d})$ in Theorem~\ref{Sec:3:Thm:Successive-Refinement} of Section~\ref{Sec:3}. However, before giving this bound, it is useful to formally define the rate-distortion function $R(\mbf{d})$  (fig.~\ref{Sec:1:Fig:mKHB}) and then review Heegard and Berger's functional $R_0(\mbf{d})$.

\subsection{Rate Distortion with Side-Information at $t$-decoders}\label{Sec:2:B}

The rate-distortion function $R(\mbf{d})$ for the problem shown in Figure~\ref{Sec:1:Fig:mKHB} can be efficiently recovered from $\set{R}(\mbf{d})$ by restricting the code rate on channels $2$ through $t$ to be zero.

\medskip

\begin{definition}\label{Sec:2:Def:Heegard-Berger}
The rate-distortion function for lossy source coding with side-information at $t$-decoders (fig.~\ref{Sec:1:Fig:mKHB}) is defined by
\begin{equation*}
R(\mbf{d}) \triangleq \min \big\{ r \in \mathbb{R}_+ : (r,0,0,\cdots,0) \in \set{R}(\mbf{d}) \big\}\ ,
\end{equation*}
where the indicated minimum exists because $\set{R}(\mbf{d})$ is closed and bounded from below.
\end{definition}

\medskip

It should be noted that Definition~\ref{Sec:2:Def:Heegard-Berger} technically permits the use of codes with \textit{asymptotically-vanishing rates} on channels $2$ through $t$. That is, the $\mbf{d}$-admissibility of rates approaching $R(\mbf{d})$ from above can be proved using a sequence of codes where $\epsilon_i \rightarrow 0$ and $\kappa^{(n_{\epsilon_i})}_l \rightarrow 0$ for all $l \in [2,t]$. Such codes, however, are not permitted in the single-channel rate-distortion problem (fig.~\ref{Sec:1:Fig:mKHB}); we can only use codes with $\kappa^{(n)}_l = 0$ for all $l \in [2,t]$. Despite this subtle difference, Definition~\ref{Sec:2:Def:Heegard-Berger} is equivalent to the definition used in~\cite{Heegard-Nov-1985-A} because any message transmitted on channels $2$ through $t$ can be transferred\footnote{In general, it is difficult to prove the equivalence of \textit{asymptotically-vanishing rates} and \textit{zero-capacity channels} (i.e. ``deleting the channel'') without such a rate-transfer argument. See, for example, ~\cite{Vellambi-Mar-2010-C}.} to channel $1$ (see Proposition~\ref{Sec:2:Pro:Sum-Incremental}).

As mentioned in Section~\ref{Sec:2:A}, $\set{R}(\mbf{d})$ depends on the successive-refinement decoding order. This dependence, of course, is not shared by $R(\mbf{d})$. Indeed, the aforementioned rate-transfer argument can be used show that the decoding order (used to define $\set{R}(\mbf{d})$ in Definition~\ref{Sec:2:Def:Heegard-Berger}) can be interchanged with any other decoding order without altering $R(\mbf{d})$.

Using the time-sharing principle, it can be shown that $R(\mbf{d})$ is convex on $\mathbb{R}_+^t$. This convexity ensures that $R(\mbf{d})$
is continuous on the interior of $\mathbb{R}_+^t$~\cite[Thm. 10.1]{Rockafellar-1997-B}. Moreover, it can also be verified that $R(\mbf{d})$ is continuous whenever $d_l = 0$ for some $l \in [t]$; see, for example, ~\cite[Pg. 2]{Wyner-Jan-1976-A}.

\medskip

\begin{proposition}
The rate-distortion function $R(\mbf{d})$ is continuous, non-increasing (i.e., $R(\mbf{d}) \leq R(\tilde{\bf{d}})$ when $d_l \geq \tilde{d}_l$ for all $l\in[t]$) and convex on $\mathbb{R}_+^t$.
\end{proposition}

\medskip

The following proposition for lossless reconstructions can be obtained as an extension to the Slepian-Wolf Theorem~\cite[Thm. 2]{Slepian-Jul-1973-A}, a variant of a more general result by Sgarro~\cite[Thm. 2]{Sgarro-Mar-1977-A}, or a special case of Bakshi and Effros~\cite[Thm. 1]{Bakshi-Jul-2008-C}.

\medskip

\begin{proposition}\label{Sec:2:Pro:Slepian-Wolf}
If, for every $l \in [t]$, $\Hat{\set{X}}_l = \set{X}$ and $\delta_l$ satisfies
\begin{equation*}
\delta_l(x,x) = 0 \text{ and}
\end{equation*}
\begin{equation*}
\delta_l(x,\Hat{x}) > 0,\quad  x \neq \Hat{x}\ ,
\end{equation*}
then
\begin{equation*}
R(0,0,\ldots,0) = \max_{l \in [t]} H(X|Y_l)\ .
\end{equation*}
\end{proposition}

\medskip

To review Heegard and Berger's work on $R(\mbf{d})$ for generic distortion tuples, we first need to define $(2^t-1)$-auxiliary random variables -- one for every non-empty subset of decoders. For this purpose, arrange the non-empty subsets of $[t]$ into a list $\set{S}_1,\set{S}_2,\ldots,\set{S}_{2^t-1}$ (the ordering is not important). For each $j\in[2^t-1]$, let $\set{U}_{\set{S}_j}$ be a finite alphabet. Define $\set{U}^* \triangleq \set{U}_{\set{S}_1} \times \set{U}_{\set{S}_2} \times \cdots \times \set{U}_{\set{S}_{2^t-1}}$.
Let $\set{P}$ denote all those pmfs $p$ on $\set{U}^* \times \set{X} \times \set{Y}^*$ whose $(\set{X} \times \set{Y}^*)$-marginal is equal to the source distribution $q$:
\begin{align*}
p(x,y_1,\ldots,y_t) &\triangleq
\sum_{(u_1,u_2\ldots,u_{2^t-1}) \in \set{U}^{*}}  p(u_1,u_2\ldots,u_{2^t-1},x,y_1,\ldots,y_t)\\
&=q(x,y_1,y_2,\ldots,y_t)\  .
\end{align*}
Each $p \in \set{P}$ specifies a joint pmf for $(2^t - 1)$-auxiliary random variables. We denote these variables by $U_{\set{S}_1}$, $U_{\set{S}_2}$, $\ldots$, $U_{\set{S}_{2^t-1}}$, where $U_{\set{S}_j}$ takes values from $\set{U}_{\set{S}_j}$. Let $\set{A}$ $\triangleq$ $\{U_{\set{S}_1}$, $U_{\set{S}_2}$, $\ldots$, $U_{\set{S}_{2^t-1}}\}$, and let
\begin{align*}
\set{A}^\supset_{\S_j}    &\triangleq \Big\{U_{\set{S}_k} \in \set{A} :\ \set{S}_k \supset \S_j \Big\}
\end{align*}
denote those auxiliary random variables associated with supersets of $\set{S}_j$.

Let $\set{P}(\mbf{d})$ denote the set of all $p \in \set{P}$ for which the following two properties are satisfied:
\begin{enumerate}
\item[\textbf{(P1)}] $p$ factors to form the Markov chain:
\begin{equation*}
\big(U_{\S_1},U_{\S_2},\ldots,U_{\S_{2^t-1}}\big) \minuso X \minuso \big(Y_1,Y_2,\ldots,Y_t\big)\ [p]\ ; \text{ and}
\end{equation*}
\item[\textbf{(P2)}] for every decoder $l \in [t]$ there exists a function $\Hat{X}_l(Y_l,U_{\{l\}},\set{A}^\supset_{\{l\}})$ with
\begin{equation*}
\mathbb{E}_p \delta_l \Big(X,\Hat{X}_l\big(Y_l,U_{\{l\}},\set{A}^\supset_{\{l\}}\big)\Big) \leq d_l\ .
\end{equation*}
\end{enumerate}

Heegard and Berger claimed~\cite[Thm. 2]{Heegard-Nov-1985-A} that the functional
\begin{equation}\label{Sec:2:Eqn:HB-Upper-Bound}
R_{0}(\mbf{d}) = \min_{p \in \set{P}(\mbf{d})} \sum_{j=1}^{2^t-1} \max_{l\in\set{S}_j} I_p\big(X;U_{\set{S}_j}\big|\set{A}^\supset_{\set{S}_j},Y_l\big)
\end{equation}
is an upper bound for $R(\mbf{d})$ for all finite alphabets $\set{U}_{\set{S}_1}$, $\set{U}_{\set{S}_2}$, $\ldots$, $\set{U}_{\set{S}_{2^t-1}}$ such that $\set{P}(\mbf{d})$ is non-empty. In the next two examples, we confirm that $R_{0}(\mbf{d})$ is an upper bound for $R(\mbf{d})$ when there is one or two decoders $(t = 1 \text{ or } 2)$; however, in the third example we show that $R_{0}(\mbf{d})$ is \textit{not} an upper bound for $R(\mbf{d})$ when there is three or more decoders ($t \geq 3$).

For brevity, we drop set notation for each auxiliary random in the following three examples. For example, we write $U_1$, $U_{12}$ and $U_{123}$ in place of $U_{\{1\}}$, $U_{\{1,2\}}$ and $U_{\{1,2,3\}}$, respectively.

\medskip

\begin{example}
If $t=1$, then \eqref{Sec:2:Eqn:HB-Upper-Bound} reduces to
\begin{align}
\notag R_0(d_1) &= \min_{p \in \set{P}(d_1)} I_p\big(X;U_1\big|Y_1\big) \\
\label{Sec:2:Eqn:HB-WZ}   &= \min_{p \in \set{P}(d_1)} \Big\{I_p\big(X;U_1\big) - I_p\big(U_1;Y_1\big)\Big\}\ ,
\end{align}
where the equality in~\eqref{Sec:2:Eqn:HB-WZ} follows from the chain rule for mutual information and the Markov chain $U_1 \minuso X \minuso Y_1\ [q]$. If the cardinality of $\set{U}_{\set{S}_1}$ is limited to $|\set{U}_{\set{S}_1}| \leq |\set{X}| + 1$, then the right hand side of~\eqref{Sec:2:Eqn:HB-WZ} reduces to the Wyner-Ziv formula~\eqref{Sec:1:Eqn:Wyner-Ziv-1}.   \hfill $\square$
\end{example}

\medskip

\begin{example}
If $t=2$, then \eqref{Sec:2:Eqn:HB-Upper-Bound} reduces to
\begin{align}
\label{Sec:2:Eqn:2-decoder}
R_0(d_1,d_2) &= \min_{p \in \set{P}(d_1,d_2)} \Big\{ \max_{l \in \{1,2\}} I_p\big(X;U_{12}\big|Y_l\big) + I_p\big(X;U_{1}\big|Y_1,U_{12}\big) + I_p\big(X;U_{2}\big|Y_2,U_{12}\big) \Big\}\ .
\end{align}
One may invoke the Support Lemma~\cite[Pg. 310]{Csiszar-1981-B} to show that imposing the cardinality constraints $|\set{U}_{\{1,2\}}| \leq |\set{X}| + 5$, $|\set{U}_{\{1\}}| \leq |\set{X}|\ |\set{U}_{{\{1,2\}}}| + 1$, and $|\set{U}_{\{2\}}| \leq |\set{X}|\ |\set{U}_{{\{1,2\}}}| + 1$, does not alter the minimization in~\eqref{Sec:2:Eqn:2-decoder}. It can be shown, see Theorem~\ref{Sec:4:Thm:HB}, that $R_0(d_1,d_2) \geq R(d_1,d_2)$. \hfill $\square$
\end{example}

\medskip

\begin{example}\label{Sec:2:Counterexample}
If $t = 3$ and $|\set{Y}_{1}|$ $=$ $|\set{Y}_{2}|$ $=$ $|\set{Y}_{3}|$ $=$ $1$, then \eqref{Sec:2:Eqn:HB-Upper-Bound} reduces to
\begin{align}
\notag R_0(d_1,d_2,d_3) = \min_{p \in \set{P}(d_1,d_2,d_3)} \Big\{ & I_p\big(X;U_{123}\big)
+ I_p\big(X;U_{12}\big|U_{123}\big) + I_p\big(X;U_{13}\big|U_{123}\big)\\
\notag &\quad + I_p\big(X;U_{23}\big|U_{123}\big) + I_p\big(X;U_{1}\big|U_{12},U_{13},U_{123}\big)\\
\label{Sec:2:Eqn:HB-CE-1} &\quad + I_p\big(X;U_{2}\big|U_{12},U_{23},U_{123}\big) + I_p\big(X;U_{3}\big|U_{13},U_{23},U_{123}\big)\Big\}\ .
\end{align}
Suppose that $\set{X}$ $=$ $\Hat{\set{X}}_1$ $=$ $\Hat{\set{X}}_2$ $=$ $\Hat{\set{X}}_3$ $=$ $\{0,1,2\}$, and let $X$ be uniform on $\set{X}$. Finally, set
\begin{equation}\label{Sec:3:Eqn:Hamming}
\delta_l(x,\Hat{x}) = \left\{
                       \begin{array}{ll}
                         0, & \hbox{if } x = \Hat{x}  \\
                         1, & \hbox{otherwise,}
                       \end{array}
                     \right.
\end{equation}
for $l = 1,2,3$ and require that $d_1 = d_2 = d_3 = 0$.

We now choose the following auxiliary random variables. Set
\begin{subequations}
\begin{equation}
\set{U}_{\{1,2\}} = \set{U}_{\{1,3\}} = \set{U}_{\{2,3\}} = \{0,1,2\}\ ,\ \text{and}
\end{equation}
\begin{equation}\label{Sec:2:BE:Assumption-1}
\big|\set{U}_{\{1\}}\big| = \big|\set{U}_{\{2\}}\big| = \big|\set{U}_{\{3\}}\big| = \big|\set{U}_{\{1,2,3\}}\big| = 1\ .
\end{equation}
Let $C$ be independent of $X$ and uniform on $\{0,1,2\}$. Using modulo-3 arithmetic, choose
\begin{equation}\label{Sec:2:BE:Assumption-2}
\aux_{12} = C, \quad \aux_{13} = X + C,\  \text{ and}\quad \aux_{23} = X + 2C\ .
\end{equation}
\end{subequations}
Note, $X$ can be written as a function of any pair of $U_{12}$, $U_{13}$ and $U_{23}$, and the Markov chain $(U_1$, $U_2$, $U_3$, $U_{12}$, $U_{13}$, $U_{23}$, $U_{123})$ $\minuso$ $X$ $\minuso$ $(Y_1$, $Y_2$, $Y_3)$ is trivially satisfied. It follows that these auxiliary random variables are defined by some $p' \in \set{P}(0,0,0)$.

From~\eqref{Sec:2:BE:Assumption-1}, it follows that~\eqref{Sec:2:Eqn:HB-CE-1} is bound from above by
\begin{equation}\label{Sec:2:BE:rate-2}
R_{0}(0,0,0) \leq  I_{p'}\left(X; \aux_{12}\right) + I_{p'}\left(X; \aux_{13}\right) + I_{p'}\left(X;\aux_{23}\right)\ .
\end{equation}
Furthermore, every mutual information term on the right hand side of~\eqref{Sec:2:BE:rate-2} is zero from~\eqref{Sec:2:BE:Assumption-2}. Since $R_0(0,0,0)$ is non-negative, it follows that $R_0(0,0,0) = 0$; however, from Proposition~\ref{Sec:2:Pro:Slepian-Wolf} we have that $R(0,0,0) = H(X) > 0$. This counterexample demonstrates that $R_{0}(\mbf{d})$ is not an upper bound for $R(\mbf{d})$. \hfill $\square$
\end{example}

\medskip

It appears that this counterexample does not invalidate any results in the rate-distortion literature. In particular, those papers that cite~\cite[Thm. 3]{Heegard-Nov-1985-A} are either concerned with the special case of $2$ decoders or stochastically degraded side-information. See, for example, ~\cite{Kaspi-Nov-1994-A,Steinberg-Aug-2004-A,Tian-Oct-2006-C,Tian-Dec-2008-A,Tian-Aug-2007-A}. The case of stochastically degraded side-information is discussed in the next section.

When $t = 3$, we can force~\eqref{Sec:2:Eqn:HB-Upper-Bound} to become an upper bound for $R(d_1,d_2,d_3)$ by modifying the set $\set{P}(d_1,d_2,d_3)$ on which the minimization takes place. Namely, if we define
\begin{equation}\label{Sec:2:Eqn:Markov-Constraints-2}
\set{P}^*(d_1,d_2,d_3) \triangleq \Bigg\{ p \in \set{P}(d_1,d_2,d_3) :
\begin{array}{l}
U_{13} \minuso (X,U_{123}) \minuso U_{12}\ [p] \\
U_{23} \minuso (X,U_{123}) \minuso (U_{12},U_{13})\ [p]
\end{array}\Bigg\}\ ,
\end{equation}
then it can be shown that
\begin{equation*}
R^*_0(d_1,d_2,d_3) \triangleq \min_{p \in \set{P}^*(d_1,d_2,d_3)} \sum_{j=1}^{7} \max_{l\in\set{S}_j} I_p\big(X;U_{\set{S}_j}\big|\set{A}^\supset_{\set{S}_j},Y_l\big)
\end{equation*}
is an upper bound for $R(d_1,d_2,d_3)$. The additional Markov chains in~\eqref{Sec:2:Eqn:Markov-Constraints-2} are sufficient to verify, via classical random coding techniques, the admissibility of rates approaching $R_{0}^*(d_1,d_2,d_3)$ from above. In general, this approach can be extended to $t \geq 3$ decoders by carefully choosing appropriate Markov chains for each of the $(2^t-1)$-auxiliary random variables\footnote{In Section~\ref{Sec:4}, we will take a slightly more general approach wherein the mutual information terms in~\eqref{Sec:2:Eqn:HB-Upper-Bound} -- rather than the minimization set $\set{P}(\mbf{d})$ -- are modified to produce an upper bound for $R(\mbf{d})$. We would like to thank Dr. Chao Tian as well as an anonymous reviewer for suggesting this more general approach.}. For example, if $U_{\set{S}_j}$ is chosen to be degenerate (constant) whenever $\set{S}_j$ is not of the form $[l,t]$ for some $l \in [t]$, then one obtains appropriate Markov chains and a valid upper bound for $R(\mbf{d})$. In fact, this particular choice of auxiliary random variables is optimal when the side-information is stochastically degraded.

\subsection{Rate-Distortion with Degraded Side-Information}\label{Sec:2:C}

The side-information, as defined by $q$, is said to be \textit{degraded} if $X \minuso Y_t \minuso Y_{t-1} \minuso \cdots \minuso Y_1\ [q]$ forms a Markov chain. The side-information $q$ is said to be \textit{stochastically degraded} if there exists a pmf $q'$ on $\set{X} \times \set{Y}^*$ where $X \minuso Y_t \minuso Y_{t-1} \minuso \cdots \minuso Y_1\ [q']$ forms a Markov chain and $q(x,y_l) = q'(x,y_l)$ for every $(x,y_l) \in \set{X} \times \set{Y}_l$ and $l \in [t]$. If $\set{R}(\mbf{d})[q]$ and $\set{R}(\mbf{d})[q']$ are the respective $\mbf{d}$-admissible rate regions for $q$ and $q'$, then this condition and Proposition~\ref{Sec:2:Pro:Marginal-Property} ensures that $\set{R}(\mbf{d})[q] = \set{R}(\mbf{d})[q']$. Thus, it is sufficient to consider degraded side-information.

When the side-information is degraded, $R(\mbf{d})$ can be characterised using $t$ auxiliary random variables. These variables are $U_{[1,t]}$, $U_{[2,t]}$, $\ldots$, $U_{\{t\}}$, and the corresponding subsets of decoders are $[1,t]$, $[2,t]$, $\ldots$, $\{t\}$. To formally define these variables using the notation of Section~\ref{Sec:2:B}, choose $|\set{U}_{\set{S}_j}|=1$ whenever $\set{S}_j \neq [l,t]$ for some $l \in [t]$, and let $\set{P}_{deg}(\mbf{d})$ denote the resultant set of $p \in \set{P}$ that satisfy properties \textbf{(P1)} and \textbf{(P2)}.

\medskip

\begin{proposition}\label{Sec:2:Pro:HB-Lower-Bound-Degraded}
If $X \minuso Y_t \minuso Y_{t-1} \minuso \cdots \minuso Y_1\ [q]$ forms a Markov chain, then
\begin{equation}
\label{Sec:2:Eqn:HB-Degraded-1}
R(\mbf{d}) = \min_{p \in \set{P}_{deg}(\mbf{d})} \sum_{l=1}^t
I_p\big(X;U_{[l,t]}\big|Y_l,U_{[1,t]},U_{[2,t]},\ldots,U_{[l-1,t]}\big)\ ,
\end{equation}
where the cardinality of each set $\set{U}_{[l,t]}$ is bound by
\begin{equation*}
\big|\set{U}_{[l,t]}\big| \leq \big|\set{X}\big| \prod_{l'=1}^{l-1}
\big|\set{U}_{[l',t]}\big| - 1 + t - l + \frac{(t-l+1)(t-l+2)}{2} \ .
\end{equation*}
\end{proposition}

\medskip

The converse theorem for this result can be found on~\cite[Pgs. 733-734]{Heegard-Nov-1985-A}. Note, however, that the use of $R_0(\mbf{d})$ in~\cite[Thm. 3]{Heegard-Nov-1985-A} is incorrect. For example, the side-information used in Example~\ref{Sec:2:Exa:Counterexample} is trivially degraded.

Finally, we note that the Markov chain $X \minuso Y_t \minuso Y_{t-1} \minuso \cdots \minuso Y_1\ [q]$ appears to be essential for the converse theorem~\cite[Pgs. 733-734]{Heegard-Nov-1985-A}. In contrast, the coding theorem that proves the admissibility of rates approaching~\eqref{Sec:2:Eqn:HB-Degraded-1} is less dependent on this assumption. Indeed, this Markov chain can be disregarded provided there is an appropriate increase in rate. For example, the functional
\begin{equation*}
\min_{p\in\set{P}_{deg}(\mbf{d})} \sum_{l=1}^t
\max_{l' \in [l,t]} I_p\big(X;U_{[l,t]}\big|Y_{l'},U_{[1,t]},\ldots,U_{[l-1,t]}\big)
\end{equation*}
is an upper bound for $R(\mbf{d})$. We will extend this idea in the next section to give an inner bound for $\set{R}(\mbf{d})$.

\section{Main Results for $\set{R}(\mbf{d})$}\label{Sec:3}

\subsection{An Inner Bound for $\set{R}(\mbf{d})$}\label{Sec:3:A}
We now present a new inner bound for $\set{R}(\mbf{d})$. This bound will require an auxiliary random variable for each non-empty subset of decoders. For this purpose, arrange the non-empty subsets of $[t]$ into an ordered list $v = \set{S}_1,\set{S}_2,\ldots,\set{S}_{2^t-1}$ with decreasing cardinality. That is, $|\set{S}_j| \geq |\set{S}_k|$ whenever $j \leq k$. Let $\set{V}$ denote the set of all such lists.

Fix $v \in  \set{V}$. Let $\set{U}_{\set{S}_1}$, $\set{U}_{\set{S}_2}$, $\ldots$, $\set{U}_{\set{S}_{2^t-1}}$ be finite alphabets and define $\set{U}_v^* \triangleq \set{U}_{\set{S}_1}$ $\times$ $\set{U}_{\set{S}_2}$ $\times$ $\cdots$ $\times$ $\set{U}_{\set{S}_{2^t-1}}$. Let $\set{P}_v$ denote the set of all distributions on $\set{U}^*_v \times \set{X} \times \set{Y}^*$ whose $(\set{X} \times \set{Y}^*)$-marginal is equal to $q$; that is, $p\big(x,y_1,\ldots,y_y\big) = q\big(x,y_1,\ldots,y_y\big)$.

As before, each $p \in \set{P}_v$ specifies a joint distribution for $(2^t-1)$-auxiliary random variables. We denote these variables by $U_{\set{S}_j}$, $j=1,2,\ldots,2^t-1$, where $U_{\set{S}_j}$ takes values from $\set{U}_{\set{S}_j}$. Let $\set{A} \triangleq \{U_{\set{S}_1},\ U_{\set{S}_2},\ \ldots, U_{\set{S}_{2^t-1}}\}$, and define
\begin{align*}
\set{A}^{-}_{\set{S}_j} &\triangleq \Big\{U_{\set{S}_i} \in \set{A}\ :\ i < j,\ \set{S}_i \nsupseteq \set{S}_j  \Big\} \text{ and}\\
\set{A}^{\supset}_{\set{S}_j} &\triangleq \Big\{ U_{\set{S}_i} \in \set{A}\ :\ \set{S}_i \supset \set{S}_j \Big\}\ .
\end{align*}
We note that the union of $\set{A}^{-}_{\set{S}_j}$ and $\set{A}^{\supset}_{\set{S}_j}$ is the set of all those auxiliary random variables associated with subsets that appear before $\set{S}_j$ in $v$. Let us further define
\begin{align*}
\set{A}^{+}_{\set{S}_j} &\triangleq \Big\{U_{\set{S}_k} \in \set{A}\ :\ k > j,\ \set{S}_k \cap \set{S}_j \neq \emptyset \Big\} \ , \\
\set{A}^{\dag}_{\set{S}_j} &\triangleq \left\{ U_{\set{S}_i} \in \set{A}^-_{\set{S}_j} :
  \begin{array}{ll}
    \exists U_{\set{S}_k} \in \set{A}^+_{\set{S}_j}, \\
    \set{S}_i \cap \set{S}_k \neq \emptyset
  \end{array}\right\}\ \text{ and}\\
\set{A}^\ddag_{\set{S}_j,l} &\triangleq \Big\{U_{\set{S}_i} \in \set{A}^\dag_{\set{S}_j}\ :\ \set{S}_i\ni l \Big\}\ \text{ when } l \in \set{S}_j\ .
\end{align*}
Finally, let $\set{P}_{v}(\mbf{d})$ denote the set of all $p \in \set{P}_v$ satisfying properties $(\textbf{P1})$ and $(\textbf{P2})$ from Section~\eqref{Sec:2:B}.

Our inner bound for $\set{R}(\mbf{d})$ will be built using the following functional. For each subset $\set{S}_j \subseteq [t]$ and $l \in [t]$ such that $\set{S}_j\cap[l]\neq\emptyset$, let
\begin{equation}
\label{Sec:3:Eqn:Phi}
\Phi_p\big(\set{S}_j,l\big) \triangleq I_p\big(X,\set{A}^\dag_{\set{S}_j};U_{\set{S}_j}\big|\set{A}^\supset_{\set{S}_j}\big)
 -\min_{l'\in\set{S}_j \cap [l]}I_p\big(U_{\set{S}_j};\set{A}^\ddag_{\set{S}_j,l'},Y_{l'}\big|\set{A}^\supset_{\set{S}_j}\big)\ .
\end{equation}
Finally, for each $p \in \set{P}_{v}(\mbf{d})$, define\footnote{One can invoke the Support Lemma~\cite[Pg.310]{Csiszar-1981-B} to upper bound the cardinality of each set $\set{U}_{\set{S}_j}$. Note, these bounds will depend on the particular choice of list $v$.}
\begin{equation*}
\set{R}_{p,v}(\mbf{d})
\triangleq \left\{ \mbf{r} \in \mathbb{R}_+^t\ :\sum_{i=1}^l r_i
\geq
\sum_{\tiny
\begin{array}{c}
  \set{S}_j \subseteq [t], \\
  \set{S}_j \cap [l] \neq \emptyset
\end{array}
}
\Phi_p\big(\set{S}_j,l\big),\ \forall l \in [t]\right\}\ ,
\end{equation*}
and let
\begin{equation*}
\set{R}_{in}(\mbf{d}) \triangleq \text{co} \left(
\bigcup_{v\in\set{V}} \bigcup_{p\in\set{P}_v(\mbf{d})} \set{R}_{p,v}(\mbf{d})\right)\ ,
\end{equation*}
where $\text{co}(\cdot)$ denotes the closure of the convex hull.

\medskip

\begin{theorem}\label{Sec:3:Thm:Successive-Refinement}
If $\mbf{d} \in \mathbb{R}_+^t$, then every rate tuple within $\set{R}_{in}(\mbf{d})$ is $\mbf{d}$-admissible; that is,
\begin{equation*}
\set{R}_{in}(\mbf{d}) \subseteq \set{R}(\mbf{d})\ .
\end{equation*}
\end{theorem}

\medskip

Our proof of this result is given in Appendix~\ref{App:1}.

\subsection{Stochastically Degraded Side-Information}\label{Sec:3:B}

Assuming that the side-information is stochastically degraded, Tian and Diggavi gave a single-letter characterisation of $\set{R}(\mbf{d})$ in~\cite[Thm. 1]{Tian-Aug-2007-A} (see Proposition~\ref{Sec:1:Pro:Degraded-Side-Information}). We now show that the forward (coding) part of this result can be obtained as a special case of Theorem~\ref{Sec:3:Thm:Successive-Refinement}.

We can assume that $X$ $\minuso$ $Y_t$ $\minuso$ $Y_{t-1}$ $\minuso$ $\cdots$ $\minuso$ $Y_1$ $[q]$ forms a Markov chain. Recall $\set{P}_{deg}(\mbf{d})$ from Section~\ref{Sec:2:C}. Each $p \in \set{P}_{deg}$ specifies a joint distribution for $t$ non-degenerate auxiliary random variables. These variables are $U_{[1,t]}$, $U_{[2,t]}$, $\ldots$, $U_{\{t\}}$ and the associated subsets are $[1,t]$, $[2,t]$, $\ldots$, $\{t\}$, respectively. We can ignore the degenerate random variables in $\set{A}$, so that for all $l \in [1,t]$ we have
\begin{subequations}
\begin{equation}\label{Sec:3:Eqn:Degraded-SI-ARV-Supset}
\set{A}^\supset_{[l,t]} = \Big\{U_{[1,t]},U_{[2,t]},\ldots,U_{[l-1,t]}\Big\}\ ,
\end{equation}
\begin{equation}\label{Sec:3:Eqn:Degraded-SI-ARV-Dag}
\set{A}^\dag_{[l,t]} = \emptyset\quad \text{ and}
\end{equation}
\begin{equation}\label{Sec:3:Eqn:Degraded-SI-ARV-DDag}
\set{A}^\ddag_{[l,t],l'} = \emptyset\quad \forall  l'\in[l,t].
\end{equation}
\end{subequations}
On combining the Markov chain $(U_{[1,t]}$, $U_{[2,t]},$ $\ldots,$ $U_{\{t\}})$ $\minuso$ $X$ $\minuso$ $(Y_1,$ $Y_2,$ $\ldots,$ $Y_t)$ $[p]$ with the Markov chain $X \minuso Y_t \minuso Y_{t-1} \minuso \cdots \minuso Y_1$ $[p]$, we obtain the following Markov chains:
\begin{equation}\label{Sec:3:Eqn:MC-1}
U_{[l,t]} \minuso \Big(\set{A}^{\supset}_{[l,t]},Y_{l'}\Big) \minuso Y_l\ [p],\quad \forall\ l' \in [l,t]\ .
\end{equation}
On substituting~\eqref{Sec:3:Eqn:Degraded-SI-ARV-Supset}, \eqref{Sec:3:Eqn:Degraded-SI-ARV-Dag} and~\eqref{Sec:3:Eqn:Degraded-SI-ARV-DDag} into~\eqref{Sec:3:Eqn:Phi}, we obtain
\begin{equation}\label{Sec:3:Eqn:Phi-Degraded}
\Phi_p([l,t],j) = I_p\Big(X;U_{[l,t]}\Big|\set{A}^\supset_{[l,t]}\Big)
- \min_{l' \in [l,j]} I_p\Big(U_{[l,t]};Y_{l'}\Big|\set{A}^\supset_{[l,t]} \Big)\ .
\end{equation}
The second term on the right hand side of~\eqref{Sec:3:Eqn:Phi-Degraded} can be rewritten as
\begin{align}
\notag
\min_{l' \in [l,j]} I_p\Big(U_{[l,t]};Y_{l'}\Big|\set{A}^\supset_{[l,t]}\Big)
&=H_p\Big(U_{[l,t]}\Big|\set{A}^\supset_{[l,t]}\Big)-
\max_{l' \in [l,j]} H_p\Big(U_{[l,t]}\Big|\set{A}^\supset_{[l,t]},Y_{l'}\Big)\\
\label{Sec:3:Eqn:Phi-Degraded-2}
&=H_p\Big(U_{[l,t]}\Big|\set{A}^\supset_{[l,t]}\Big)-
\max_{l' \in [l,j]} H_p\Big(U_{[l,t]}\Big|\set{A}^\supset_{[l,t]},Y_{l'},Y_l\Big)\\
\label{Sec:3:Eqn:Phi-Degraded-3}
&=H_p\Big(U_{[l,t]}\Big|\set{A}^\supset_{[l,t]}\Big)- H\Big(U_{[l,t]}\Big|\set{A}^\supset_{[l,t]},Y_l\Big)\\
\label{Sec:3:Eqn:Phi-Degraded-4}
&=I_p\Big(U_{[l,t]};Y_l\Big|\set{A}^\supset_{[l,t]}\Big)\ ,
\end{align}
where~\eqref{Sec:3:Eqn:Phi-Degraded-2} follows from the Markov chain~\eqref{Sec:3:Eqn:MC-1}, and~\eqref{Sec:3:Eqn:Phi-Degraded-3} follows since
\begin{equation*}
H_p\Big(U_{[l,t]}\Big|\set{A}^\supset_{[l,t]},Y_l\Big) \geq H_p\Big(U_{[l,t]}\Big|\set{A}^\supset_{[l,t]},Y_{l'},Y_l\Big)\ , \quad \forall l' \in [l,j]\ .
\end{equation*}
On combining~\eqref{Sec:3:Eqn:Phi-Degraded} and~\eqref{Sec:3:Eqn:Phi-Degraded-4}, we get
\begin{align}
\notag
\Phi_p([l,t],j)
&=I_p\Big(X;U_{[l,t]}\Big|\set{A}^\supset_{[l,t]}\Big)
- I_p\Big(U_{[l,t]};Y_{l}\Big|\set{A}^\supset_{[l,t]} \Big)\\
\label{Sec:3:Eqn:Phi-Degraded-5}
&= I_p\Big(X,\set{A}^\supset_{[l,t]};U_{[l,t]}\Big)
- I_p\Big(U_{[l,t]};Y_{l},\set{A}^\supset_{[l,t]}\Big).
\end{align}
From~\eqref{Sec:3:Eqn:Degraded-SI-ARV-Supset} and since
$U_{[l,t]} \minuso (X,\set{A}^{\supset}_{[l,t]}) \minuso Y_l\ [p]$
forms a Markov chain, \eqref{Sec:3:Eqn:Phi-Degraded-5} further simplifies to
\begin{equation}\label{Sec:3:Eqn:Phi-Degraded-6}
\Phi_p([l,t],j) = I_p\Big(X;U_{[l,t]}\Big|U_{[1,t]},U_{[2,t]},\ldots,U_{[l-1,t]},Y_l\Big)\ .
\end{equation}
Finally, substituting~\eqref{Sec:3:Eqn:Phi-Degraded-6} into the definition of $\set{R}_{p,v}(\mbf{d})$ proves the $\mbf{d}$-admissibility of every rate tuple $\mbf{r} \in \mathbb{R}_+^t$ for which there exists some $p \in \set{P}_{deg}(\mbf{d})$ with
\begin{equation*}
\sum_{i=1}^j r_i \geq \sum_{l=1}^j I_p\Big(X;U_{[l,t]}\Big|U_{[1,t]},U_{[2,t]},\ldots,U_{[l-1,t]},Y_l\Big)\ ,
\end{equation*}
for $j = 1,2,\ldots,t$.

\subsection{Side-Information Scalable Source Coding}\label{Sec:3:C}
If $t=2$ and the side-information is degraded ($X \minuso Y_2 \minuso Y_1 [q]$), then an optimal compression strategy should satisfy the distortion constrains of decoder $2$ after the distortion constraints of decoder $1$ have been satisfied. See, for example, Section~\ref{Sec:2:C}. However, this ordering may not be optimal when the side-information is not degraded. This observation led Tian and Diggavi in~\cite[Thm. 1]{Tian-Dec-2008-A} (see Proposition~\ref{Sec:1:Pro:SI-Scalable}) to propose and study the side-information scalable source coding problem. In the context of this paper, this problem is a special case of the successive-refinement problem where $X \minuso Y_1 \minuso Y_2$ $[q]$ is assumed to form a Markov chain. We now show that this result can be obtained as a special case of Theorem~\ref{Sec:3:Thm:Successive-Refinement}.

Choose the list $v$ as follows: $\S_1 = \{1,2\}$, $\S_2 = \{1\}$ and $\S_3 = \{2\}$. For each $p \in \set{P}_v(d_1,d_2)$, we have the chains $X \minuso Y_1 \minuso Y_2$ $[p]$ and $(U_{12},U_1,U_2) \minuso X \minuso (Y_1,Y_2)$, therefore \eqref{Sec:3:Eqn:Phi} simplifies to
\begin{align*}
\Phi_p(\{1,2\},1) &= I_p(X;U_{12}) - I_p(U_{12};Y_1)\\
                  &= I_p(X;U_{12}|Y_1)\\
\Phi_p(\{1,2\},2) &= I_p(X;U_{12}) - \min_{l'\in\{1,2\}} I_p(U_{12};Y_{l'})\\
                  &= I_p(X;U_{12}) - I_p(U_{12};Y_{2})\\
                  &= I_p(X;U_{12}|Y_2)\\
\Phi_p(\{1\},1)   &= I_p(X;U_1|U_{12}) - I_p(U_1;Y_1|U_{12})\\
                  &= I_p(X;U_1|U_{12},Y_1)\\
\Phi_p(\{1\},2)   &= I_p(X;U_1|U_{12},Y_1)\\
\Phi_p(\{2\},2)   &= I_p(X;U_2|U_{12}) - I_p(U_2;Y_2|U_{12})\\
                  &= I_p(X;U_2|U_{12},Y_2)\ .
\end{align*}
On substituting these equalities into the definition of $\set{R}_{v,p}(d_1,d_2)$, it can been seen from Theorem~\ref{Sec:3:Thm:Successive-Refinement} that any rate pair $(r_1,r_2)$ satisfying
\begin{align*}
r_1 &\geq \Phi_p(\{1,2\},1) + \Phi_p(\{1\},1)\\
&= I_p(X;U_{12}|Y_1) + I_p(X;U_1|U_{12},Y_1)\\
&= I_p(X;U_1,U_{12}|Y_1)\ ,
\end{align*}
and
\begin{align*}
r_1 + r_2 &\geq \Phi_p(\{1,2\},2) + \Phi_p(\{1\},2) + \Phi_p(\{2\},2)\\
&= I_p(X;U_{12}|Y_2) + I_p(X;U_1|U_{12},Y_1) + I_p(X;U_2|U_{12},Y_2) \\
&= I_p(X;U_2,U_{12}|Y_2) + I_p(X;U_1|U_{12},Y_1)
\end{align*}
for some $p \in \set{P}_v(d_1,d_2)$ is $\mbf{d}$-admissible. This condition matches the desired inner bound~\cite[Thm. 1]{Tian-Dec-2008-A} (Proposition~\ref{Sec:1:Pro:SI-Scalable}).

\section{Main Results for the Wyner-Ziv Problem with $t$-Decoders}\label{Sec:4}

\subsection{An Upper Bound for $R(\mbf{d})$}
Recall Figure~\ref{Sec:1:Fig:mKHB} and the rate-distortion function $R(\mbf{d})$.

\begin{theorem}\label{Sec:4:Thm:HB}
\begin{equation}\label{Sec:4:Eqn:HB-Upper-Bound}
R(\mbf{d}) \leq \min_{
\begin{tiny}
\begin{array}{c}
        v\in\set{V},\\
p \in \set{P}_v(\mbf{d})
\end{array}
\end{tiny}
}
\sum_{j =1}^{2^t-1} \left[I_p\big(X,\set{A}^\dag_{\set{S}_j};U_{\set{S}_j}\big|\set{A}^\supset_{\set{S}_j}\big) - \min_{l' \in \set{S}_j} I_p\big(U_{\set{S}_j};\set{A}^\ddag_{\set{S}_j,l'},Y_{l'}|\set{A}^\supset_{\set{S}_j}\big)\right]\ .
\end{equation}
\end{theorem}

\medskip

We note the following special cases where this upper bound known to be tight. For one decoder, the right hand side of~\eqref{Sec:4:Eqn:HB-Upper-Bound} gives the Wyner-Ziv formula~\eqref{Sec:1:Eqn:Wyner-Ziv-1}. For $t$-decoders and degraded side-information, the right hand side of~\eqref{Sec:4:Eqn:HB-Upper-Bound} is equal to the right hand side of~\eqref{Sec:2:Eqn:HB-Degraded-1}. (Set $|\set{U}_{\set{S}_j}|=1$ whenever $\set{S}_j \neq [l,t]$ for some $l \in [t]$, and following the reasoning given in Section~\ref{Sec:3:B}.) In fact, this upper bound is tight whenever $X$ $\minuso$ $Y_{\alpha_1}$ $\minuso$ $Y_{\alpha_2}$ $\minuso$ $\cdots$ $\minuso$ $Y_{\alpha_t}$, where $\alpha_l$, $l = 1,2,\ldots,t$ each take unique values from $[t]$ (see Remark~\ref{Sec:5:Rem:DeterministicDistortionMeasures}). Most importantly, however, this bound avoids those problems suffered by $R_0(\mbf{d})$ in Example~\ref{Sec:2:Counterexample}.

\subsection{Proof of Theorem~\ref{Sec:4:Thm:HB}}

The following lemma will be useful for the proof of Theorem~\ref{Sec:4:Thm:HB}.

\medskip

\begin{lemma}\label{Sec:4:Lem:Phi}
Suppose $p \in \set{P}_v(\mbf{d})$, and recall the functional $\Phi_p(\set{S}_j,l)$ defined in~\eqref{Sec:3:Eqn:Phi}. For every $\set{S}_j \subseteq [t]$ and $l,l' \in [t]$ such that $\S_j \cap [l] \neq \emptyset$ and $\S_j \cap [l'] \neq \emptyset$, we have:
\begin{enumerate}
\item[(i)] $\Phi_p(\set{S}_j,l) \leq \Phi_p(\set{S}_j,l')$ when $l' > l$, and
\item[(ii)] $\Phi_p(\set{S}_j,l) \geq 0$.
\end{enumerate}
\end{lemma}

\medskip

\begin{proof}
The fact that $\Phi_p(\set{S}_j,l) \leq \Phi_p(\set{S}_j,l')$ follows because $\set{S}_j \cap [l] \subseteq \set{S}_j \cap [l']$. To see that $\Phi_p(\set{S}_j,l) \geq 0$, consider the following. Let
\begin{equation*}
\tilde{l} \triangleq \underset{i \in \set{S}_j \cap [l]}{\operatorname{argmin}}\ I_p\big(U_{\set{S}_j};\set{A}^\ddag_{\set{S}_j,i},Y_i\big|\set{A}^{\supset}_{\set{S}_j}\big)\ ,
\end{equation*}
then
\begin{align}
\notag
\Phi_p\big(\set{S}_j,l\big) &\equiv I_p\big(X,\set{A}^\dag_{\set{S}_j};U_{\set{S}_j}\big|\set{A}^{\supset}_{\set{S}_j}\big)
-I_p\big(U_{\set{S}_j};\set{A}^\ddag_{\set{S}_j,\tilde{l}},Y_{\tilde{l}}\big|\set{A}^{\supset}_{\set{S}_j}\big)\\
\label{Sec:3:Eqn:Phi-1}
&= I_p\big(X,\set{A}^\dag_{\set{S}_j},Y_{\tilde{l}};U_{\set{S}_j}\big|\set{A}^{\supset}_{\set{S}_j}\big)
-I_p\big(U_{\set{S}_j};\set{A}^\ddag_{\set{S}_j,\tilde{l}},Y_{\tilde{l}}\big|\set{A}^{\supset}_{\set{S}_j}\big)\\
\label{Sec:3:Eqn:Phi-2}
&=I_p\big(X,\set{A}^{\supset}_{\set{S}_j},\set{A}^\dag_{\set{S}_j},Y_{\tilde{l}};U_{\set{S}_j}\big)
-I_p\big(U_{\set{S}_j};\set{A}^{\supset}_{\set{S}_j},\set{A}^\ddag_{\set{S}_j,\tilde{l}},Y_{\tilde{l}}\big)\\
\label{Sec:3:Eqn:Phi-3}
&\geq 0\ ,
\end{align}
where~\eqref{Sec:3:Eqn:Phi-1} follows because $Y_{\tilde{l}} \minuso (X,\set{A}^{\supset}_{\set{S}_j},\set{A}^{\dag}_{\set{S}_j}) \minuso U_{\set{S}_j}$ $[p]$ forms a Markov chain, \eqref{Sec:3:Eqn:Phi-2} follows from the chain rule for mutual information, and~\eqref{Sec:3:Eqn:Phi-3} follows from $\set{A}^\dag_{\set{S}_j} \supseteq \set{A}^\ddag_{\set{S}_j,\tilde{l}}$.
\end{proof}

\medskip

We now prove Theorem~\ref{Sec:4:Thm:HB}. First, note that the minimum on the right hand side of~\eqref{Sec:4:Eqn:HB-Upper-Bound} exists. Suppose that $v$ and $p$ achieve this minimum, and choose any $r \in \reals_+$ such that
\begin{equation}\label{Sec:4:Eqn:HB-Bound-Proof-3}
r \geq
\sum_{j =1}^{2^t-1} \left[I_p\big(X,\set{A}^\dag_{\set{S}_j};U_{\set{S}_j}\big|\set{A}^\supset_{\set{S}_j}\big) - \min_{l' \in \set{S}_j} I_p\big(U_{\set{S}_j};\set{A}^\ddag_{\set{S}_j,l'},Y_{l'}|\set{A}^\supset_{\set{S}_j}\big)\right]\ .
\end{equation}
In the following, we prove the $\mbf{d}$-admissibility of $r$ using Theorem~\ref{Sec:3:Thm:Successive-Refinement}.

Consider the successive refinement problem shown in Figure~\ref{Sec:1:Fig:tSR}, the corresponding $\mbf{d}$-admissible rate region $\set{R}(\mbf{d})$ (defined in Section~\ref{Sec:2:A}), and the inner bound $\set{R}_{in}(\mbf{d})$ given in Theorem~\ref{Sec:3:Thm:Successive-Refinement}. In particular, consider the region $\set{R}_{p,v}(\mbf{d})$, where $v$ and $p$ achieve the aforementioned minimum. Define the $t$-tuple $\tilde{\mbf{r}} \triangleq (r,0,0,\ldots,0)$. It is clear that $r \geq R(\mbf{d})$ iff $\tilde{\mbf{r}}\in \set{R}(\mbf{d})$, therefore the result will follow if it can be shown that $\tilde{\mbf{r}}\in \set{R}_{p,v}(\mbf{d})$.

For every $l \in [t]$, we have
\begin{align}
\notag
\sum_{i=1}^l \tilde{r}_i &\geq \sum_{j =1}^{2^t-1} \left[I_p\big(X,\set{A}^\dag_{\set{S}_j};U_{\set{S}_j}\big|\set{A}^\supset_{\set{S}_j}\big) - \min_{l' \in \set{S}_j} I_p\big(U_{\set{S}_j};\set{A}^\ddag_{\set{S}_j,l'},Y_{l'}|\set{A}^\supset_{\set{S}_j}\big)\right]\\
\label{Sec:4:Eqn:HB-Bound-Proof-1a}
&=
\sum_{\S_j \subseteq [t]}\Phi_p\big(\set{S}_j,t\big)\\
\label{Sec:4:Eqn:HB-Bound-Proof-1}
&\geq
\sum_{\tiny
\begin{array}{c}
  \set{S}_j \subseteq [t], \\
  \set{S}_j \cap [l] \neq \emptyset
\end{array}
}
\Phi_p\big(\set{S}_j,t\big)
\end{align}

\begin{align}
\label{Sec:4:Eqn:HB-Bound-Proof-2}
&\geq
\sum_{\tiny
\begin{array}{c}
  \set{S}_j \subseteq [t], \\
  \set{S}_j \cap [l] \neq \emptyset
\end{array}
}
\Phi_p\big(\set{S}_j,l\big)\ ,
\end{align}
where~\eqref{Sec:4:Eqn:HB-Bound-Proof-1a} follows from~\eqref{Sec:3:Eqn:Phi}, and Lemma~\ref{Sec:4:Lem:Phi} gives~\eqref{Sec:4:Eqn:HB-Bound-Proof-1} and~\eqref{Sec:4:Eqn:HB-Bound-Proof-2}. From Theorem~\ref{Sec:3:Thm:Successive-Refinement} we have that $\tilde{\mbf{r}} \in \set{R}_{v,p}(\mbf{d})$ and $\tilde{\mbf{r}} \in \set{R}(\mbf{d})$, therefore $r \geq R(\mbf{d})$. \hfill $\square$

\begin{remark}
Theorem~\ref{Sec:4:Thm:HB} is a consequence of the inner bound $\set{R}_{in}(\mbf{d})$ given in Theorem~\ref{Sec:3:Thm:Successive-Refinement}. Like $\set{R}(\mbf{d})$, $\set{R}_{in}(\mbf{d})$ depends on the successive-refinement decoding order: if we interchange the decoders (keeping the same side-information and distortion constraints at each decoder), then the resulting inner bound $\set{R}_{in}(\mbf{d})$ will change. One might, therefore, be inspired to pursue a stronger version of Theorem~\ref{Sec:4:Thm:HB} wherein the choice of successive-refinement order is optimized. Note, however, that the proof of Theorem~\ref{Sec:4:Thm:HB} requires only the bound for $r_1 + r_2 + \cdots + r_t$ in $\set{R}_{p,v}(\mbf{d})$, and this bound is independent of the successive-refinement decoding order.
\end{remark}

\section{Lossless Source Coding with Private Messages}\label{Sec:5}

In Proposition~\ref{Sec:2:Pro:Slepian-Wolf}, we reviewed a broadcast problem wherein $\mbf{X}$ is reconstructed losslessly at every decoder. This lossless problem can be easily solved as a variant of existing work by Slepian and Wolf~\cite[Thm. 2]{Slepian-Jul-1973-A};
Sgarro~\cite[Thm. 2]{Sgarro-Mar-1977-A}; or Bakshi and Effros~\cite[Thm. 1]{Bakshi-Jul-2008-C}. In this section, we consider a more complex scenario wherein each decoder is required to decode one part of $\mbf{X}$ losslessly.

\begin{figure}[t]
  \centering
\includegraphics[width=0.75\columnwidth]{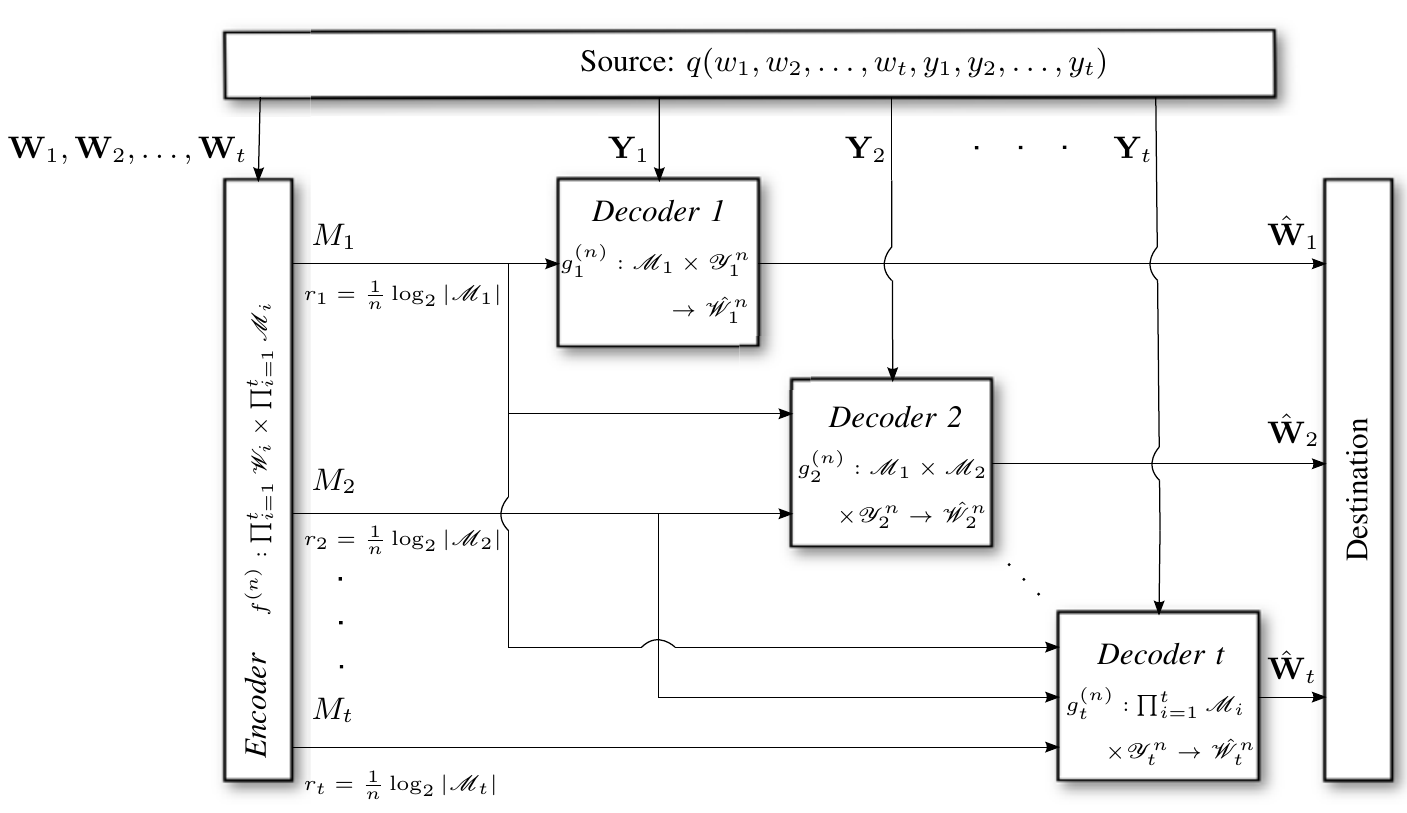}
  \caption{Lossless Source Coding with Private Messages. The encoder compresses $(\mbf{W}_1,\mbf{W}_2,\ldots,\mbf{W}_t)$ to $(M_1,M_2,\ldots,M_t)$. It is required that decoder $l$ uses $M_1$ through $M_l$ together with $\mbf{Y}_l$ to produce a lossless replica $\Hat{\mbf{W}_l}$ of $\mbf{W}_l$. In Theorem~\ref{Sec:4:Thm:Lossless}, we give an explicit characterisation of the $\set{R}(0,0,\ldots,0)$ for degraded side-information $(W_1,$ $W_2,$ $\ldots,$ $W_t)$ $\minuso$ $Y_t$ $\minuso$ $Y_{t-1}$ $\minuso$ $\cdots$ $\minuso$ $Y_1$.}
  \label{Sec:4:Fig:Lossless}
\end{figure}

Let $\set{W}_1$, $\set{W}_2$, $\ldots$, $\set{W}_t$ be finite alphabets, and consider the problem shown Figure~\ref{Sec:4:Fig:Lossless}. In the nomenclature of previous sections, set $\set{X}$ $\triangleq$ $\set{W}_1$ $\times$ $\set{W}_2$ $\times$ $\cdots$ $\times$ $\set{W}_t$, $X \triangleq (W_1,$ $W_2,$ $\ldots,$ $W_t)$, and let $(\mbf{W}_1$, $\mbf{W}_2$, $\ldots$, $\mbf{W}_t$, $\mbf{Y}_1$, $\mbf{W}_2$, $\ldots$, $\mbf{W}_t)$ be drawn iid according to $q(w_1,w_2,\ldots,w_t,y_1,y_2,\ldots,y_t)$. It is required that decoder $l$ reconstructs $\mbf{W}_l$ with vanishing probability of symbol error. To this end, set $\Hat{\set{X}}_l \triangleq \set{W}_l$ and define the average symbol error probability at decoder $l$ to be
\begin{align*}
P^l_{e} &\triangleq \frac{1}{n} \sum_{i=1}^n P^l_{e,i}
\end{align*}
where $P^l_{e,i} \triangleq \mathbb{E} [\delta_l(W_{1,i},W_{2,i},\ldots,W_{t,i},\Hat{W}_{l,i})]$,
\begin{equation}\label{Sec:5:Eqn:Distortion-Measure}
\delta_l\big(w_1,w_2,\ldots,w_t,\Hat{w}_l\big) \triangleq \left\{
    \begin{array}{ll}
      0, & \hbox{ if } w_l = \Hat{w}_l \\
      1, & \hbox{ otherwise,}
    \end{array}
  \right.
\end{equation}
defines the probability of error for the $i^{\text{th}}$-symbol.

A computable characterisation of $\set{R}(0,0,\ldots,0)$ has yet to be found. A direct application of Theorem~\ref{Sec:3:Thm:Successive-Refinement} yields an inner bound for $\set{R}(0,0,\ldots,0)$; however, it is not clear if this bound is tight. The next theorem shows that this bound is tight when the side-information is degraded. Although this result is a special case of Proposition~\ref{Sec:1:Pro:Degraded-Side-Information}, we state it here in an explicit form -- without auxiliary random variables -- to highlight the generality of this problem.

\medskip

\begin{theorem}\label{Sec:4:Thm:Lossless}
If $(W_1,$ $W_2,$ $\ldots,$ $W_t)$ $\minuso$ $Y_t$ $\minuso$ $Y_{t-1}$ $\minuso$ $\cdots$ $\minuso$ $Y_1$ $[q]$ and $\delta_l$ is given by~\eqref{Sec:5:Eqn:Distortion-Measure}, then
\begin{equation*}
\set{R}(0,0,\ldots,0) = \Bigg\{ \mbf{r} \in \reals_+^t : \sum_{k=1}^l r_k \geq \sum_{k=1}^l H\big(W_k\big|W_1,W_2,\ldots,W_{k-1},Y_k\big)\Bigg\}
\end{equation*}
\end{theorem}

\medskip

The lossless one-channel version of Theorem~\ref{Sec:4:Thm:Lossless} follows immediately.

\medskip

\begin{corollary}\label{Sec:4:Cor:Lossless}
If $(W_1,$ $W_2,$ $\ldots,$ $W_t)$ $\minuso$ $Y_t$ $\minuso$ $Y_{t-1}$ $\minuso$ $\cdots$ $\minuso$ $Y_1$ $[q]$ and $\delta_l$ is given by~\eqref{Sec:5:Eqn:Distortion-Measure}, then
\begin{equation*}
R(0,0,\ldots,0) = \sum_{l=1}^t H\big(W_l\big|W_1,W_2,\ldots,W_{l-1},Y_l\big)\ .
\end{equation*}
\end{corollary}

\medskip

\begin{remark}\label{Sec:5:Rem:DeterministicDistortionMeasures}
The lossless problems considered in this section are equivalent to the concept of deterministic distortion measures~\cite{Fu-July-2002-A,Tian-Dec-2008-A}, wherein certain functions $\{\phi_i(X)\}$ of the source $X$ are to be reconstructed with vanishing symbol error probability at the receivers. If $t = 2$, $Z_1 = \phi_1(X)$ is to be reconstructed at receiver $1$, $Z_2 = \phi_2(X)$ is to be reconstructed at receiver $2$, and the side-information is reversibly degraded (i.e. $X \minuso Y_1 \minuso Y_2$ $[q]$ forms a Markov chain), then Tian and Diggavi have shown that~\cite[Cor. 4]{Tian-Dec-2008-A}
\begin{equation*}
R(0,0) = H(Z_2|Y_2) + H(Z_1|Y_1,Z_2)\ .
\end{equation*}
This result is consistent with Corollary~\ref{Sec:4:Cor:Lossless} in the following sense. The achievability of Corollary~\ref{Sec:4:Cor:Lossless} follows from Theorem~\ref{Sec:4:Thm:HB}
by setting $U_{\S_j} = W_l$ whenever $\S_j = [l,t]$ for some $l \in [t]$ and $U_{\S_j} = \text{constant}$ otherwise. The bound in Theorem~\ref{Sec:4:Thm:HB} is equal to the rate-distortion function $R(\mbf{d})$ for every order of degraded side-information. For example, suppose that $X$ $\minuso$ $Y_{\alpha_t}$ $\minuso$ $Y_{\alpha_{t-1}}$ $\minuso$ $\cdots$ $\minuso$ $Y_{\alpha_1}$ $[q]$ forms a Markov chain, where $\alpha_l$, $l = 1,2,\ldots,t$ each take unique values from $[t]$. This markov condition is simply a relabelling of the degradedness considered in Section~\ref{Sec:2:C}, so it is appropriate to choose the $t$ non-trivial auxiliary random variables to be $U_{[\alpha_1,\alpha_t]}$, $U_{[\alpha_2,\alpha_t]}$, $\ldots$, $U_{\{\alpha_t\}}$, where $[\alpha_i,\alpha_t] = \{\alpha_i,\alpha_{i+1},\ldots,\alpha_{t}\}$. Thus, we can set $U_{[\alpha_i,\alpha_t]} = W_{\alpha_i}$ to restate Corollary~\ref{Sec:4:Cor:Lossless} for an arbitrary order of degraded side-information.

Tian and Diggavi also characterise the successive-refinement region $\set{R}(0,0)$ in~\cite[Thm. 4]{Tian-Dec-2008-A} for $t = 2$ and reversibly degraded side-information. This result is not captured by Theorem~\ref{Sec:4:Thm:Lossless}, and it would be interesting to see if a similar result can be obtained for $t$-receivers and arbitrary ordering of degraded side-information.
\end{remark}

\medskip

\begin{proof}
The forward (coding) part follows from  by setting $U_{l} = W_l$ in Proposition~\ref{Sec:1:Pro:Degraded-Side-Information}. The converse theorem requires some work and is given below. For brevity, we use the following notation: $M_{\leq l} \triangleq \{M_1,M_2,\ldots,M_l\}$, $\mbf{W}_{\leq l} \triangleq \{\mbf{W}_1,\mbf{W}_2,\ldots,\mbf{W}_l\}$ and $\mbf{Y}_{\leq l} \triangleq \{\mbf{Y}_1,\mbf{Y}_2,\ldots,\mbf{Y}_l\}$. By definition, we have
\begin{align}
\label{Sec:4:Eqn:Con-Proof-Lossless-1}
\sum_{k=1}^l \big(r_k + \epsilon\big) &\geq
\frac{1}{n}\sum_{k=1}^l \log_2|\set{M}_l| \\
&\geq \frac{1}{n} H\big(M_{\leq l}\big) \\
\label{Sec:4:Eqn:Con-Proof-Lossless-2}
&\geq \frac{1}{n} I\big(\mbf{W}_{\leq l},\mbf{Y}_{\leq l};M_{\leq l}\big)\\
\label{Sec:4:Eqn:Con-Proof-Lossless-3}
&= \frac{1}{n} \sum_{k=1}^l I\big(\mbf{W}_k,\mbf{Y}_k;M_{\leq l}\big|\mbf{W}_{\leq k-1},\mbf{Y}_{\leq k-1}\big)\\
\label{Sec:4:Eqn:Con-Proof-Lossless-4}
&\geq \frac{1}{n} \sum_{k=1}^l I\big(\mbf{W}_k;M_{\leq l}\big|\mbf{W}_{\leq k-1},\mbf{Y}_{\leq k}\big)\\
\label{Sec:4:Eqn:Con-Proof-Lossless-6}
&= \frac{1}{n} \sum_{k=1}^l \Big[H\big(\mbf{W}_k\big|\mbf{W}_{\leq k-1},\mbf{Y}_{\leq k}\big)  - H\big(\mbf{W}_k\big|M_{\leq l} ,\mbf{W}_{\leq k-1},\mbf{Y}_{\leq k}\big)\Big]\\
\notag
&= \frac{1}{n} \sum_{k=1}^l \sum_{i=1}^n
\Big[H\big(W_{k,i}\big|\mbf{W}_{\leq k-1},\mbf{Y}_{\leq k},W_{k,1},W_{k,2},\ldots,W_{k,i-1}\big)\\
\label{Sec:4:Eqn:Con-Proof-Lossless-7}
&\qquad \qquad \qquad \qquad -H\big(W_{k,i}\big|M_{\leq l},\mbf{W}_{\leq k-1},\mbf{Y}_{\leq k},W_{k,1},W_{k,2},\ldots,W_{k,i-1}\big)\Big]\\
\label{Sec:4:Eqn:Con-Proof-Lossless-9}
&\geq \sum_{k=1}^l
H\big(W_{k}\big|W_{1},W_{2},\ldots,W_{k-1},Y_{k}\big)
-\frac{1}{n} \sum_{k=1}^l \sum_{i=1}^n H\big(W_{k,i}\big|\Hat{W}_{k,i}\big)\\
\label{Sec:4:Eqn:Con-Proof-Lossless-10}
&\geq \sum_{k=1}^l
H\big(W_{k}\big|W_{1},W_{2},\ldots,W_{k-1},Y_{k}\big)
-\frac{1}{n} \sum_{k=1}^l \sum_{i=1}^n \Big[h\big(P^k_{e,i}\big)+P^k_{e,i}\log_2|\set{W}_k|\Big]\\
\label{Sec:4:Eqn:Con-Proof-Lossless-11}
&\geq \sum_{k=1}^l
H\big(W_{k}\big|W_{1},W_{2},\ldots,W_{k-1},Y_{k}\big)
-\sum_{k=1}^l  \Big[h\big(P^k_{e}\big)+P^k_{e}\log_2|\set{W}_k|\Big]\\
\label{Sec:4:Eqn:Con-Proof-Lossless-12}
& \geq \sum_{l=1}^t H\big( W_l \big| W_1, W_2,\ldots,W_{l-1},Y_l\big) - \Big[l\ h(\epsilon) + \epsilon \log_2|\set{W}_1|\ |\set{W}_2|\ \cdots |\set{W}_l|\Big]\ ,
\end{align}
where~\eqref{Sec:4:Eqn:Con-Proof-Lossless-1} through~\eqref{Sec:4:Eqn:Con-Proof-Lossless-7} follow from standard Shannon inequalities; \eqref{Sec:4:Eqn:Con-Proof-Lossless-9} follows because $(\mbf{W}_1$, $\ldots$, $\mbf{W}_t$, $\mbf{Y}_1$, $\ldots$, $\mbf{Y}_t)$ is iid, $W_{k}$ $\minuso$ $(W_{1}$, $W_{2}$, $\ldots$, $W_{k-1}$, $Y_{k})$ $\minuso$ $(Y_{1}$, $Y_{2}$, $\ldots$, $Y_{k-1})$ forms a Markov chain; conditioning reduces entropy and $\Hat{W}_{k,i}$ is a function of $M_1,M_2,\ldots,M_k$ and $\mbf{Y}_k$; \eqref{Sec:4:Eqn:Con-Proof-Lossless-10} follows from Fano's Inequality where $h(\cdot)$ is the binary entropy function~\cite{Cover-1991-B}; \eqref{Sec:4:Eqn:Con-Proof-Lossless-11} follows from the concavity of $h(\cdot)$ and Jensen's inequality; \eqref{Sec:4:Eqn:Con-Proof-Lossless-12} follows by assuming $\epsilon$ is small (i.e. $0  < P^k_e < \epsilon < 1/2$). Finally, $l\ h(\epsilon)$ $+$ $\epsilon \log_2|\set{W}_1|\ |\set{W}_2|\ \cdots |\set{W}_l|\rightarrow 0$ as $\epsilon \rightarrow 0$.
\end{proof}

\section{Conclusion}\label{Sec:6}
We studied the rate-distortion function $R(\mbf{d})$ and the rate region $\set{R}(\mbf{d})$ for the problems shown in Figures~\ref{Sec:1:Fig:mKHB} and~\ref{Sec:1:Fig:tSR}, respectively. In~\cite[Thm. 2]{Heegard-Nov-1985-A}, Heegard and Berger claimed that a certain functional, $R_0(\mbf{d})$, is an upper bound for $R(\mbf{d})$. By way of a counterexample, we demonstrated that $R_0(\mbf{d})$ is not an upper bound for $R(\mbf{d})$. In Theorem~\ref{Sec:4:Thm:HB}, we gave a new upper bound for $R(\mbf{d})$. This bound followed from a new inner bound for $\set{R}(\mbf{d})$ that we presented in Theorem~\ref{Sec:3:Thm:Successive-Refinement}. Finally, we gave an explicit characterisation of the rates needed to losslessly reconstruct private messages at each decoder (assuming degraded side-information) in Theorem~\ref{Sec:4:Thm:Lossless}.

\section*{Acknowledgements}
The authors would like to thank Dr. Chao Tian and an anonymous reviewer whose comments helped generalise Theorem~\ref{Sec:3:Thm:Successive-Refinement} from a more restricted statement into its current form.

\appendices

\section{Proof of Theorem~\ref{Sec:3:Thm:Successive-Refinement}}\label{App:1}

Fix $v\in\set{V}$ and $p\in\set{P}_v(\mbf{d})$ arbitrarily. It is sufficient to prove the $\mbf{d}$-admissibility of rate tuples within $\set{R}_{p,v}(\mbf{d})$. (The $\mbf{d}$-admissibility of tuples within $\set{R}_{in}(\mbf{d})$ follows by standard time-sharing arguments.) Our proof uses a random-coding argument that is based on the concept of $\epsilon$-letter typical sequences\footnote{We have reviewed the relevant $\epsilon$-letter typical results in Appendix~\ref{App:B} for convenience; a more detailed treatment can be found in~\cite{Kramer-2008-A}.}. This argument employs $(2^t-1)$-randomly generate codebooks; one codebook for every non-empty subset of receivers. The encoder selects a codeword from each codebook and sends some information (the bin indices of each codeword) to the decoders. Each decoder tries to recover those codewords where it is a member of the corresponding subset. To help elucidate the main ideas of the proof, we present the special case of four decoders as a series of examples in parallel to the main proof.

For notational convenience, we impose the natural ordering on the elements of each subset $\S_j$, and we let $\set{S}_j[i]$ denote the $i^{\text{th}}$-smallest element of $\set{S}_j$. For example, if $\S_j = \{1,3,5\}$, then $\S_j[1] = 1$, $\S_j[2] = 3$ and $\S_j[3] = 5$.

\subsection{Code Construction}
For each subset $\S_j$, construct an $|\S_j|$-layer nested codebook in the following manner. For each vector-valued index
$\mbf{k}_{\S_j} \triangleq ( k_{\S_j,1},k_{\S_j,2},\ldots, k_{\S_j,|\S_j|},k'_{\S_j} )$,
where
\begin{equation*}
k_{\S_j,i} = 1,2,\ldots,2^{nR_{\S_j,i}}\ , \quad i = 1,2,\ldots,|\S_j|\ ,
\end{equation*}
\begin{equation*}
k'_{\S_j} = 1,2,\ldots,2^{nR'_{\S_j}}\ ,
\end{equation*}
generate a length $n$ codeword $\mbf{u}_{\S_j}(\mbf{k}_{\S_j}) \in \set{U}^n_{\S_j}$ by selecting $n$ symbols from $\set{U}_{\S_j}$ in an iid manner using $p(u_{\set{S}_j})$ -- the $U_{\S_j}$-marginal of $p$. The values of $R_{\S_j,i}$ and $R_{\S_j}'$ will be defined shortly.

\medskip

\begin{example}[$4$-Decoders Code Construction]
Choose the list $v$ as follows: $\S_1 = \{1,2,3,4\}$, $\S_2 = \{1,2,3\}$, $\S_3 = \{1,2,4\}$, $\S_4 = \{1,3,4\}$, $\S_5 = \{2,3,4\}$, $\S_6 =  \{1,2\}$, $\S_7 =  \{1,3\}$, $\S_8 =  \{1,4\}$, $\S_9 =  \{2,3\}$, $\S_{10} =  \{2,4\}$, $\S_{11} = \{3,4\}$, $\S_{12} =  \{1\}$, $\S_{13} = \{2\}$, $\S_{14} = \{3\}$ and $\S_{15} = \{4\}$. Figure~\ref{Sec:7:Fig:Code-Layers} shows the
$3$-layer nested codebook associated with the subset $\{1,2,3\}$. In the first layer, there are $2^{nR_{123,1}}$ bins (labelled with the index $k_{123,1}$) each of which contain $2^{n(R'_{123} + R_{123,2} + R_{123,3})}$ codewords. The set of codewords inside a particular layer one bin define the second layer of the codebook. Specifically, each layer one index $k_{123,1} \in [2^{nR_{123,1}}]$ identifies $2^{nR_{123,2}}$ layer two bins. These bins are labelled with the index $k_{123,2}$, and each bin contains $2^{n(R'_{123} + R_{123,3})}$ codewords. Similarly, each pair  $k_{123,1} \in [2^{nR_{123,1}}]$ and $k_{123,2} \in [2^{nR_{123,2}}]$ identifies $2^{nR_{123,3}}$ layer  three bins. There are $2^{n(R'_{123})}$ codewords in each one of the layer three bins.

\begin{figure*}
  \centerline{\includegraphics[scale=0.8]{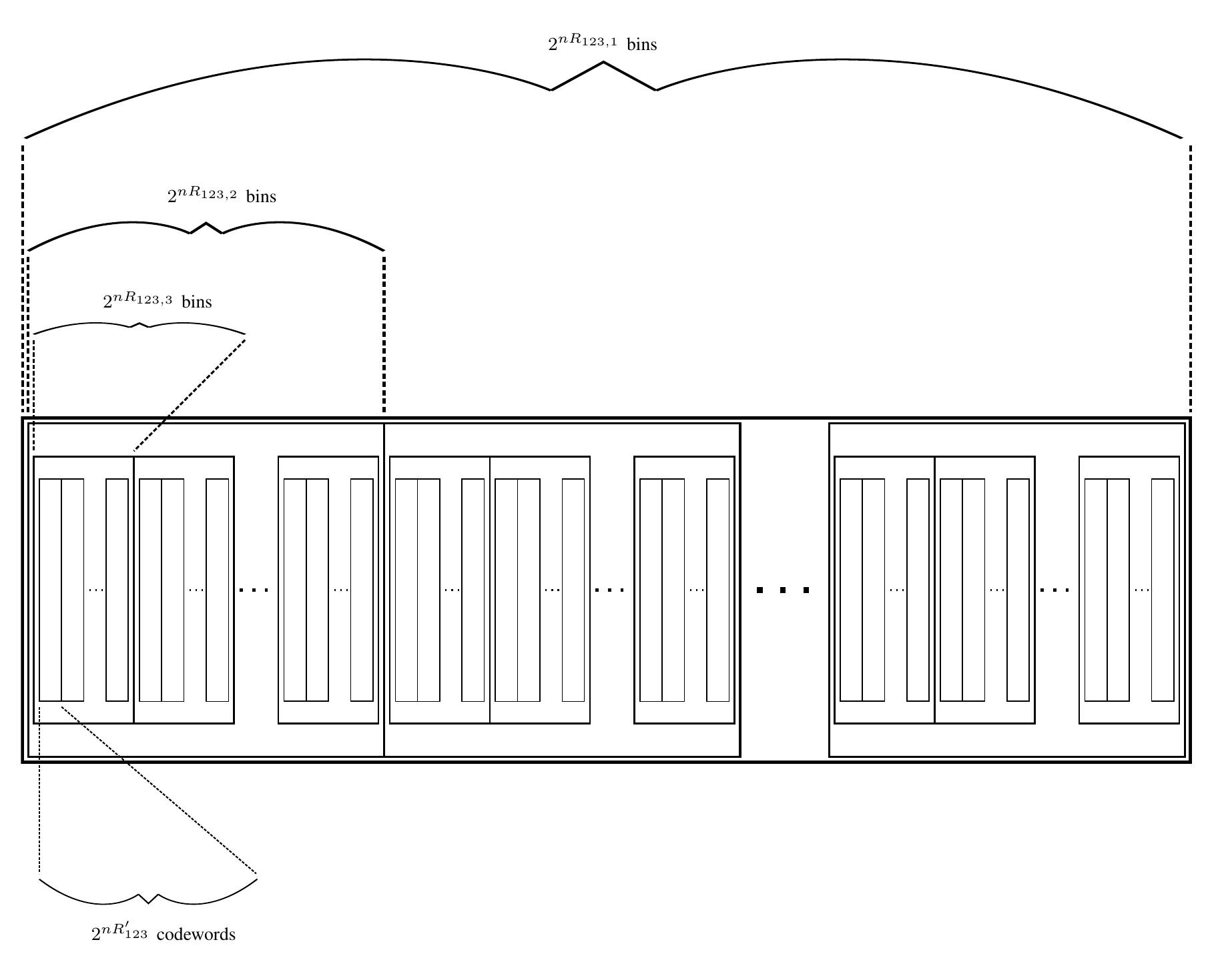}\\}
  \caption{$4$-Decoders Example: Figure shows the nested bin structure for the codebook $\S_2 = \{1,2,3\}$.}\label{Sec:7:Fig:Code-Layers}
\end{figure*}

\end{example}

\subsection{Encoding}

Encoding proceeds sequentially over $(2^t - 1)$-stages using $\epsilon$-letter typical-set encoding rules. For this purpose, choose $0< \epsilon_0 < \epsilon_1 < \cdots < \epsilon_{2^t}$ to be arbitrarily small real numbers. The encoder is given $\mbf{x} \in \set{X}^n$. At encoding stage $j$ it selects the codebook with label $\S_j$ and looks for an index vector $\k_{\S_j}$ where the corresponding codeword $\mbf{u}_{\S_j}(\k_{\S_j})$ is $\epsilon_{j}$-letter typical with
$\mbf{x}$ and
\begin{subequations}\label{Sec:Proof:Eqn:Typ-Enc-1}
\begin{align}
\label{Sec:Proof:Eqn:Typ-Enc-1a}
\mbf{u}_{\S_j}^{\supset} &\triangleq  \big\{ \mbf{u}_{\S_i}(\k_{\S_i}) : i < j,\ \S_i \supset \S_j \big\}, \text{ and}\ \\
\label{Sec:Proof:Eqn:Typ-Enc-1b}
\mbf{u}^\dag_{\S_j} &\triangleq  \big\{ \mbf{u}_{\S_i}(\k_{\S_i}) : i<j,\ \S_i \nsupseteq \S_j,\ \exists \S_{i'},\ i' > j,\ \S_{i'} \cap \S_j \neq \emptyset,\ \S_{i} \cap \S_{i'}\neq \emptyset\big\}\ .
\end{align}
\end{subequations}
If successful\footnote{If there are two-or-more such codewords, we assume that the encoder selects one codeword arbitrarily and sends the corresponding indices.}, the encoder sends the bin index $k_{\S_j,i}$ over channel $\S_j[i]$ for every $i = 1,2,\dots,|\S_j|$. If unsuccessful, the encoder sends $k_{\S_j,i}=1$ over each of these channels.

Note the correspondence between the sets $\mbf{u}_{\S_j}^{\supset}$ and $\mbf{u}^\dag_{\S_j}$ and the sets of auxiliary random variables $\set{U}_{\S_j}^{\supset}$ and $\set{U}_{\S_j}^{\dag}$, respectively. Finally, note that when $|\S_j|\geq 3$, then $\mbf{u}_{\S_j}^{\supset} \cup \mbf{u}^\dag_{\S_j} = \{ \mbf{u}_{\S_i}(\k_{\S_i}) : i < j \}$; that is, the encoder chooses $\mbf{u}_{\S_j}(\k_{\S_j})$ to be jointly typical with every codeword it has previously selected. The situation is more complex when $|\S_j|\leq 2$.

\begin{table}
\begin{footnotesize}
\centering
\begin{equation*}
\begin{array}{|c|c|c|c|}
\hline
\text{Subset } \S_j
& \mbf{u}^{\supset}_{\S_j}, \Hat{\mbf{u}}^{\supset}_{\S_j}
& \mbf{u}^{\dag}_{\S_j}
& \Hat{\mbf{u}}^{\ddag}_{\S_j,l} \\
\hline
\S_1 = \{1,2,3,4\}
& \emptyset
& \emptyset
& \emptyset \\
\hline
\S_2 = \{1,2,3\}
& \{\mbf{u}_{\S_1}\}
& \emptyset
& \emptyset \\
\hline
\S_3 = \{1,2,4\}
& \{\mbf{u}_{\S_1}\}
&\{\mbf{u}_{\S_2}\}
& \begin{array}{c}
    \Hat{\mbf{u}}^\ddag_{\S_3,1} = \{\mbf{u}_{\S_2}\}\\
    \Hat{\mbf{u}}^\ddag_{\S_3,2} = \{\mbf{u}_{\S_2}\}\\
    \Hat{\mbf{u}}^\ddag_{\S_3,4} = \emptyset
  \end{array}
\\
\hline
\S_4 = \{1,3,4\}
& \{\mbf{u}_{\S_1}\}
& \{\mbf{u}_{\S_2},\mbf{u}_{\S_3}\}
& \begin{array}{c}
    \Hat{\mbf{u}}^\ddag_{\S_4,1} = \{\mbf{u}_{\S_2},\mbf{u}_{\S_3}\} \\
    \Hat{\mbf{u}}^\ddag_{\S_4,3} = \{\mbf{u}_{\S_2}\} \\
    \Hat{\mbf{u}}^\ddag_{\S_4,4} = \{\mbf{u}_{\S_3}\}
  \end{array}\\
\hline
\S_5 = \{2,3,4\}
& \{\mbf{u}_{\S_1}\}
& \{\mbf{u}_{\S_2},\mbf{u}_{\S_3},\mbf{u}_{\S_4}\}
& \begin{array}{c}
    \Hat{\mbf{u}}^\ddag_{\S_5,2} = \{\mbf{u}_{\S_2},\mbf{u}_{\S_3}\}\\
    \Hat{\mbf{u}}^\ddag_{\S_5,3} = \{\mbf{u}_{\S_2},\mbf{u}_{\S_4}\}\\
    \Hat{\mbf{u}}^\ddag_{\S_5,4} = \{\mbf{u}_{\S_3},\mbf{u}_{\S_4}\}
  \end{array}
\\
\hline
\S_6 = \{1,2\}
& \{\mbf{u}_{\S_1},\mbf{u}_{\S_2},\mbf{u}_{\S_3}\}
& \{\mbf{u}_{\S_4},\mbf{u}_{\S_5}\}
& \begin{array}{c}
    \Hat{\mbf{u}}^\ddag_{\S_6,1} = \{\mbf{u}_{\S_4}\} \\
    \Hat{\mbf{u}}^\ddag_{\S_6,2} = \{\mbf{u}_{\S_5}\}
  \end{array}
\\
\hline
\S_7 = \{1,3\}
& \{\mbf{u}_{\S_1},\mbf{u}_{\S_2},\mbf{u}_{\S_4}\}
& \{\mbf{u}_{\S_3},\mbf{u}_{\S_5},\mbf{u}_{\S_6}\}
& \begin{array}{c}
    \Hat{\mbf{u}}^\ddag_{\S_7,1} = \{\mbf{u}_{\S_3},\mbf{u}_{\S_6}\}\\
    \Hat{\mbf{u}}^\ddag_{\S_7,3} = \{\mbf{u}_{\S_5}\}
  \end{array}
\\
\hline
\S_8 = \{1,4\}
& \{\mbf{u}_{\S_1},\mbf{u}_{\S_3},\mbf{u}_{\S_4}\}
& \{\mbf{u}_{\S_2},\mbf{u}_{\S_5},\mbf{u}_{\S_6},\mbf{u}_{\S_7}\}
& \begin{array}{c}
    \Hat{\mbf{u}}^\ddag_{\S_8,1} = \{\mbf{u}_{\S_2},\mbf{u}_{\S_6},\mbf{u}_{\S_7}\}\\
    \Hat{\mbf{u}}^\ddag_{\S_8,4} = \{\mbf{u}_{\S_5}\}
  \end{array}
\\
\hline
\S_9 = \{2,3\}
& \{\mbf{u}_{\S_1},\mbf{u}_{\S_2},\mbf{u}_{\S_5}\}
& \begin{array}{c}
    \{\mbf{u}_{\S_3},\mbf{u}_{\S_4},\mbf{u}_{\S_6},\mbf{u}_{\S_7}, \\
    \mbf{u}_{\S_8}\}
  \end{array}
& \begin{array}{c}
    \Hat{\mbf{u}}^\ddag_{\S_9,2} = \{\mbf{u}_{\S_3},\mbf{u}_{\S_6}\}\\
    \Hat{\mbf{u}}^\ddag_{\S_9,3} = \{\mbf{u}_{\S_4},\mbf{u}_{\S_7}\}
  \end{array}
\\
\hline
\S_{10} = \{2,4\}
& \{\mbf{u}_{\S_1},\mbf{u}_{\S_3},\mbf{u}_{\S_5}\}
& \begin{array}{c}
    \{\mbf{u}_{\S_2},\mbf{u}_{\S_4},\mbf{u}_{\S_6},\mbf{u}_{\S_7}, \\
    \mbf{u}_{\S_8},\mbf{u}_{\S_9}\}
  \end{array}
& \begin{array}{c}
    \Hat{\mbf{u}}^\ddag_{\S_{10},2} = \{\mbf{u}_{\S_2},\mbf{u}_{\S_6},\mbf{u}_{\S_9}\}\\
    \Hat{\mbf{u}}^\ddag_{\S_{10},4} = \{\mbf{u}_{\S_4},\mbf{u}_{\S_8}\}
  \end{array}
\\
\hline
\S_{11} = \{3,4\}
& \{\mbf{u}_{\S_1},\mbf{u}_{\S_4},\mbf{u}_{\S_5}\}
& \begin{array}{c}
    \{\mbf{u}^\ddag_{\S_2},\mbf{u}_{\S_3},\mbf{u}_{\S_7},\mbf{u}_{\S_8}, \\
    \mbf{u}^\ddag_{\S_9},\mbf{u}_{\S_{10}}\}
  \end{array}
& \begin{array}{c}
    \Hat{\mbf{u}}^\ddag_{\S_{11},3} = \{\mbf{u}_{\S_2},\mbf{u}_{\S_7},\mbf{u}_{\S_9}\}\\
    \Hat{\mbf{u}}^\ddag_{\S_{11},4} = \{\mbf{u}_{\S_3},\mbf{u}_{\S_8},\mbf{u}_{\S_{10}}\}
  \end{array}
\\
\hline
\S_{12} = \{1\}
& \begin{array}{c}
    \{\mbf{u}_{\S_1},\mbf{u}_{\S_2},\mbf{u}_{\S_3},\mbf{u}_{\S_4}, \\
    \mbf{u}_{\S_6},\mbf{u}_{\S_7},\mbf{u}_{\S_8}\}
  \end{array}
& \emptyset
& \emptyset\\
\hline
\S_{13} = \{2\}
& \begin{array}{c}
    \{\mbf{u}_{\S_1},\mbf{u}_{\S_2},\mbf{u}_{\S_3},\mbf{u}_{\S_5}, \\
    \mbf{u}_{\S_6},\mbf{u}_{\S_9},\mbf{u}_{\S_{10}}\}
  \end{array}
& \emptyset
& \emptyset\\
\hline
\S_{14} = \{3\}
& \begin{array}{c}
    \{\mbf{u}_{\S_1},\mbf{u}_{\S_2},\mbf{u}_{\S_4},\mbf{u}_{\S_5}, \\
    \mbf{u}_{\S_7},\mbf{u}_{\S_9},\mbf{u}_{\S_{11}}\}
  \end{array}
& \emptyset
& \emptyset\\
\hline
\S_{15} = \{4\}
& \begin{array}{c}
    \{\mbf{u}_{\S_1},\mbf{u}_{\S_3},\mbf{u}_{\S_4},\mbf{u}_{\S_5}, \\
    \mbf{u}_{\S_8},\mbf{u}_{\S_{10}},\mbf{u}_{\S_{11}}\}
  \end{array}
& \emptyset
& \emptyset\\
\hline
\end{array}
\end{equation*}
\end{footnotesize}
\caption{The table lists the fifteen encoding sets $\mbf{u}^{\supset}_{\S_j}$ and $\mbf{u}^{\dag}_{\S_j}$ as well as the decoding sets $\Hat{\mbf{u}}^{\supset}_{\S_j}$ and $\Hat{\mbf{u}}^{\ddag}_{\S_j}$ for the four decoder example.}
\label{Sec:Proof:Tab:1}
\end{table}

\medskip

\begin{example}[$4$-Decoders Encoding] Table~\ref{Sec:Proof:Tab:1} lists the fifteen encoding sets $\mbf{u}^{\supset}_{\S_j}$ and $\mbf{u}^{\dag}_{\S_j}$  and Figure~\ref{Sec:App:Fig:4DecodersIndexAssignement} depicts the index to channel assignments for the four decoder example. In stage $1$, the encoder considers subset $\S_1$ and looks for an index vector $\mbf{k}_{\S_1}$ such that the corresponding codeword $\mbf{u}_{\S_1}(\mbf{k}_{\S_1})$ is jointly typical with $\mbf{x}$. (The sets $\mbf{u}^{\supset}_{\S_1}$ and $\mbf{u}^{\dag}_{\S_1}$ are empty -- see Table~\ref{Sec:Proof:Tab:1}.) The resulting indices $k_{\S_1,1}$, $k_{\S_1,2}$, $k_{\S_1,3}$ and $k_{\S_1,4}$ are sent over channels $1$, $2$, $3$ and $4$, respectively. In the eleventh encoding stage, takes the codebook for $\S_{11} = \{3,4\}$ and looks for a index vector $\mbf{k}_{\S_{11}} = (k_{\S_{11},1},k_{\S_{11},2},k'_{\S_{11}})$ such that the corresponding codeword $\mbf{u}_{\S_{11}}(\mbf{k}_{\S_{11}})$ is jointly typical with $\mbf{x}$, $\mbf{u}_{\S_{1}}(\mbf{k}_{\S_{1}})$ through to $\mbf{u}_{\S_{5}}(\mbf{k}_{\S_{5}})$ and $\mbf{u}_{\S_{7}}(\mbf{k}_{\S_{7}})$ through to $\mbf{u}_{\S_{10}}(\mbf{k}_{\S_{10}})$. (Note, that this codeword need not be jointly typical with $\mbf{u}_{\S_{6}}(\mbf{k}_{\S_{6}})$.) The resulting indices $k_{\S_{11},1}$, $k_{\S_{11},2}$ are sent over channels $3$ and $4$, respectively.

\begin{figure}
  \centerline{\includegraphics[width=.75\columnwidth]{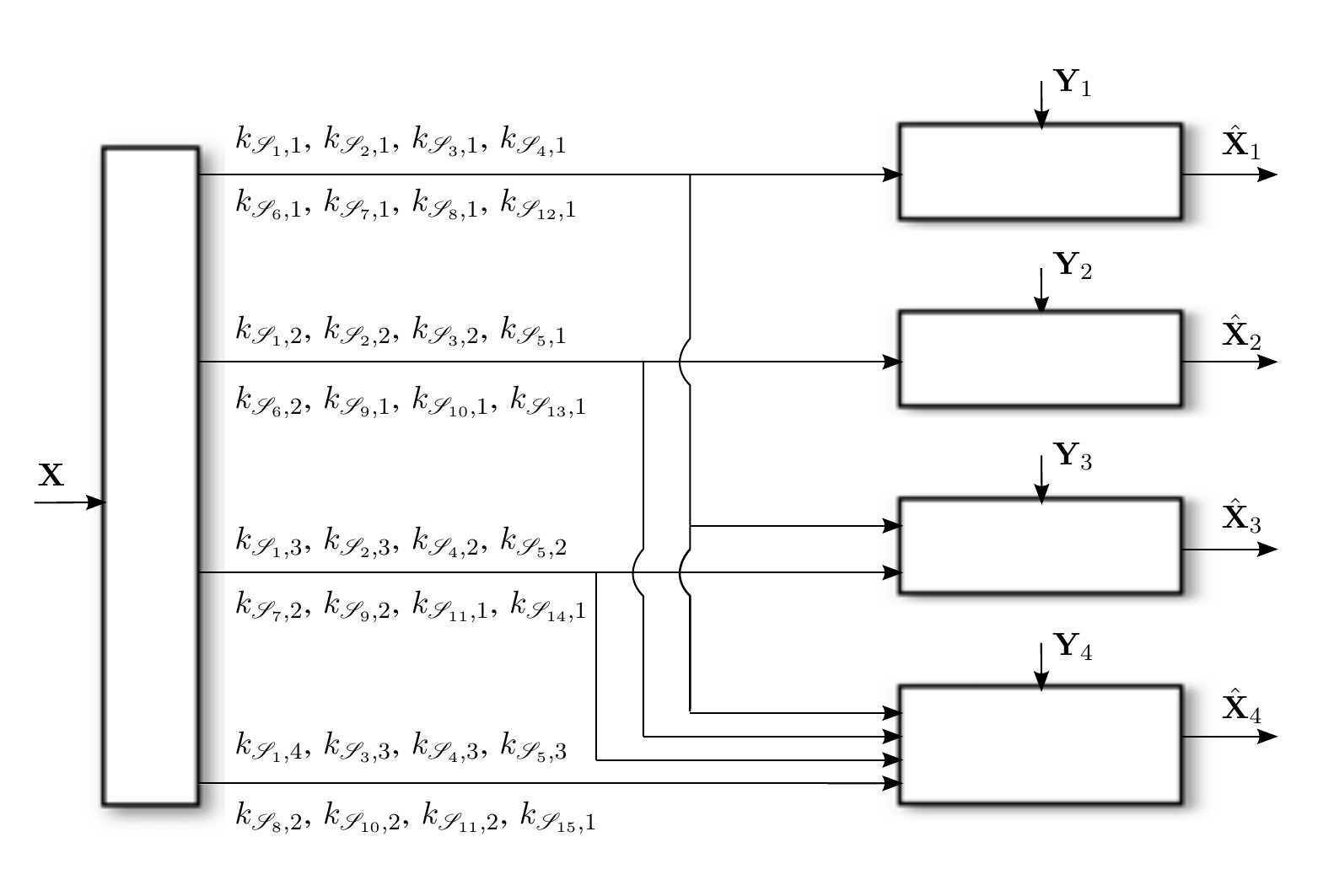}\\}
  \caption{$4$-Decoders Example: Assignment of bin indices to channels. Subsets are $\S_1 = \{1,2,3,4\}$, $\S_2 = \{1,2,3\}$, $\S_3 = \{1,2,4\}$, $\S_4 = \{1,3,4\}$, $\S_5 = \{2,3,4\}$, $\S_6 =  \{1,2\}$, $\S_7 =  \{1,3\}$, $\S_8 =  \{1,4\}$, $\S_9 =  \{2,3\}$, $\S_{10} =  \{2,4\}$, $\S_{11} = \{3,4\}$, $\S_{12} =  \{1\}$, $\S_{13} = \{2\}$, $\S_{14} = \{3\}$ and $\S_{15} = \{4\}$. Bin index $k_{\S_j,i}$ is sent over channel $\S_j[i]$.}
  \label{Sec:App:Fig:4DecodersIndexAssignement}
\end{figure}
\end{example}

\subsection{Decoding}
Consider decoder $l$. Like the encoding procedure, decoder $l$ forms its reconstruction $\Hat{\mbf{X}}_l$ of $\mbf{X}$ using $(2^t-1)$-decoding stages. Recall, this decoder recovers every bin index transmitted on channels $1$ through $l$; it does not have access to any index transmitted on channels $l + 1$ through $t$.

In stage $j$ decoder $l$ considers subset $\S_j$. If $l \notin \S_j$, then it does nothing and moves to decoding stage $j + 1$. If $l \in \set{S}_j$, then the decoder forms a reconstruction  $\mbf{u}_{\set{S}_j}(\Hat{\mbf{k}}_{\set{S}_j})$ of the codeword $\mbf{u}_{\set{S}_j}(\mbf{k}_{\set{S}_j})$, which was selected by the encoder, using the following procedure. Note, decoder $l$ will have reconstructed the following codewords in decoding stages $1$ through $j-1$:
\begin{subequations}\label{eq:decoder-1}
\begin{equation}
\Hat{\mbf{u}}_{\S_j}^{\supset} \triangleq  \big\{ \mbf{u}_{\S_i}(\Hat{\k}_{\S_i}) : i < j,\ \S_i \supset \S_j \big\}, \text{ and}
\end{equation}
\begin{equation}
\Hat{\mbf{u}}^\ddag_{\S_j,l} \triangleq  \big\{ \mbf{u}_{\S_i}(\Hat{\k}_{\S_i}) : i<j,\ \S_i \ni l,\ \S_i \nsupseteq \S_j\big\}\ .
\end{equation}
\end{subequations}
Note the correspondence between the decoding sets in~\eqref{eq:decoder-1} and the sets of auxiliary random variables $\set{A}^{\supset}_{\S_j}$ and $\set{A}^\ddag_{\S_j,l}$.

To form its reconstruction $\mbf{u}_{\set{S}_j}(\Hat{\mbf{k}}_{\set{S}_j})$, decoder $l$ takes the bin indices
\begin{equation*}
  \left\{ k_{\S_j,i} ;\  i = 1,2,\ldots,|[l] \cap \S_j| \right\}
\end{equation*}
from channels $1$ through $l$. It then looks for an index vector $\tilde{\mbf{k}}_{\S_j}$, with $\tilde{k}_{\S_j,i} = k_{\S_j,i}$ for all $i = 1,2,\ldots,|[l]\cap\S_j|$, such that the corresponding codeword $\mbf{u}(\tilde{\mbf{k}}_{\S_j})$ is $\epsilon_{j+1}$-letter typical with $\mbf{y}_l$ as well as the codewords in~\eqref{eq:decoder-1} that were decoded in the first $(j-1)$-stages:
\begin{equation}\label{eq:decoder}
\left(\Hat{\mbf{u}}_{\S_j}^{\supset},\ \Hat{\mbf{u}}^\ddag_{\S_j,l},\ \mbf{u}_{\S_j}(\tilde{\k}_{\S_j}), \mbf{y}_{l} \right) \in T_{\epsilon_{j+1}}^{(n)}(p) .
\end{equation}
Note that there are
\begin{equation*}
\exp_2\left[ {n\left(R'_{\S_j} + \sum_{i = |[l]\cap
        \S_j|+1}^{|\S_j|} R_{\S_j,i}\right)} \right]
\end{equation*}
codewords in the bin specified by the indices $\{k_{\S_j,i} : i = 1,2,\ldots,|[l] \cap \S_j|\}$. If one or more of these codewords satisfy this typicality condition, then decoder $l$ selects one arbitrarily and sets $\Hat{\k}_{\S_j} = \tilde{\k}_{\S_j}$. If there is no such codeword, it sets each of the unknown indices equal to $1$.

\medskip

\begin{example}[$4$-Decoders Decoding]
Consider the second decoder $(l = 2)$. In stage one, take $k_{\S_1,1}$ (from channel $1$) and $k_{\S_1,2}$ (from channel $2$) and look for a vector
$\tilde{\mbf{k}}_{\S_1}$ $=$ $(k_{\S_1,1}$, $k_{\S_1,2}$, $\tilde{k}_{\S_1,3},\tilde{k}'_{\S_1})$ such that the corresponding codeword $\mbf{u}_{\S_1}(\tilde{\mbf{k}}_{\S_1})$ is typical with $\mbf{y}_2$. Similarly, in stage nine take $k_{\S_9,1}$ (from channel $2$) and look for $\tilde{\mbf{k}}_{\S_9} = (k_{\S_9,1},\tilde{k}_{\S_9,2},\tilde{k}'_{\S_9})$ such that the corresponding codeword $\mbf{u}_{\S_9}(\tilde{\mbf{k}}_{\S_9})$ is jointly typical with $\mbf{y}_2$ and $\mbf{u}_{\S_1}(\Hat{\mathbf{k}}_{\S_1})$, $\mbf{u}_{\S_2}(\Hat{\mathbf{k}}_{\S_2})$, $\mbf{u}_{\S_3}(\Hat{\mathbf{k}}_{\S_3})$, $\mbf{u}_{\S_5}(\Hat{\mathbf{k}}_{\S_5})$ and $\mbf{u}_{\S_6}(\Hat{\mathbf{k}}_{\S_6})$, which were decoded during stages one through six. Finally, in stage thirteen take $k_{\S_{13},1}$ (from channel $2$) and look for $\tilde{\mbf{k}}_{\S_{13}} = (k_{\S_{12},1},\tilde{k}'_{\S_{13}})$ such that the corresponding codeword $\mbf{u}_{\S_{13}}(\tilde{\mbf{k}}_{\S_6})$ is jointly typical with $\mbf{y}_2$ and $\mbf{u}_{\S_1}(\Hat{\mathbf{k}}_{\S_1})$, $\mbf{u}_{\S_2}(\Hat{\mathbf{k}}_{\S_2})$, $\mbf{u}_{\S_3}(\Hat{\mathbf{k}}_{\S_3})$, $\mbf{u}_{\S_5}(\Hat{\mathbf{k}}_{\S_5})$,  $\mbf{u}_{\S_6}(\Hat{\mathbf{k}}_{\S_6})$, $\mbf{u}_{\S_9}(\Hat{\mathbf{k}}_{\S_9})$ and $\mbf{u}_{\S_{10}}(\Hat{\mathbf{k}}_{\S_{10}})$, which were decoded during stages one through ten.
\end{example}

\subsection{Error Analysis: Encoding}

The coding scheme is based on $\epsilon$-letter typical set encoding and decoding techniques. As such, the distortion criteria at each decoder will not be satisfied when $\left(\mbf{x}, \mbf{y}_{1}, \mbf{y}_{2}, \dots, \mbf{y}_{t}\right) \notin T^{(n)}_{\epsilon_0}(p)$. We denote this event by $E_1$.
From Lemma~\ref{App:Lem:1}, the probability of this event may be bound by
\begin{equation*}
\Pr\left[E_{1} \right] \leq \delta_1 \left(n,\epsilon_0,\mu(p)\right)\ ,
\end{equation*}
where $\delta_1 \left(n,\epsilon_0,\mu(p)\right) {\rightarrow} 0$  as $n\rightarrow\infty$.

Assume $E_1$ does not occur. Let $E_{2,\S_j}$ denote the event that the encoder fails to find an $\epsilon_j$-letter typical codeword during stage $j$ of encoding procedure given that it found an $\epsilon_i$-letter typical codeword for every stage $i\in [j-1]$. From Lemma~\ref{App:Lem:2} and the inequality $(1 - x)^t \leq e^{-tx}$ we have
\begin{align}
\notag
\Pr \left[E_{2,\S_j} \right]
&= \Bigg[ 1 -\text{Pr}\Big[\big(\mbf{u}^{\supset}_{\S_j},\mbf{u}^{\dag}_{\S_j},\mbf{U}_{\S_j}(\mbf{k}_{\S_j}),\mbf{x}\big) \in T^{(n)}_{\epsilon_{j+1}}(p)\Big]\Bigg]^{2^{n \left( R'_{\S_j} + \sum_{i=1}^{|S_j|} R_{\S_j,i}\right)}}\\
\label{Sec:7:Eqn:Error:E2-4}
& \leq
\exp \Bigg(-\big(1 - \delta_2\big)
2^{n \left( R'_{\S_j} + \sum_{i=1}^{|S_j|} R_{\S_j,i}\right)}
\cdot 2^{-n \Big(I\left(\set{A}^\supset_{\S_j},\set{A}^\dag_{\S_j},X;\aux_{\S_j}\right)
+2\epsilon_{j} H\left(\aux_{\S_j}\right)\Big)}\Bigg)
\end{align}
where we have written the function $\delta_2\left(n,\epsilon_{j-1},\epsilon_{j},\mu(p)\right)$ as $\delta_2$ for compact representation.

Let $E_2$ denote the event where a typical codeword cannot be found at
any one of the encoding stages. By the union bound we get the following upper bound for $\text{Pr}[E_2]$:
\begin{align*}
\Pr \left[E_{2}\right]
\leq
\sum_{j=1}^{2^t-1} \exp \Bigg[&-\big(1 - \delta_2\big)
2^{n \left( R'_{\S_j} + \sum_{i=1}^{|S_j|} R_{\S_j,i}\right)}
\cdot 2^{-n \Big[I\left(\set{A}^\supset_{\S_j},\set{A}^\dag_{\S_j},X;U_{\S_j}\right)
+2\epsilon_{j} H\left(U_{\S_j}\right)\Big]}\Bigg]\ .
\end{align*}
Finally, note that if
\begin{equation}\label{Sec:7:Eqn:Error:E2-5}
R'_{\S_j} + \sum_{i=1}^{|\S_j|} R_{\S_j,i}
>  I\left(\set{A}^\supset_{\S_j}, \set{A}^\dag_{\S_j},X;\aux_{\S_j}\right) + 2\epsilon_{j} H\left(\aux_{\S_j}\right)
\end{equation}
for every $j = 1,2,\ldots,2^t - 1$, then $\Pr[E_{2}]
\rightarrow 0$ as $n \rightarrow \infty$.

\subsection{Error Analysis: Decoding}

Assume $E_1$ and $E_2$ do not occur. Consider decoder $l$ and a non-trivial decoding stage $j$ where $\S_j \ni l$. Let $D_{l,\S_j}$ be the event that it cannot find a unique codeword that satisfies the typicality condition \eqref{eq:decoder} given that at every stage $i < j$ (where $\S_i \ni l$) it found a unique codeword $\mbf{u}(\Hat{\mbf{k}}_{\S_i})$ satisfying this typicality condition.

By the Markov lemma (Lemma~\ref{App:Lem:3}), the probability that the codewords $\Hat{\mbf{u}}_{\S_j}(\k_{\S_j})$, $\Hat{\mbf{u}}^{\supset}_{\S_j}$, $\mbf{u}^{\ddag}_{\S_j,l}$  are not jointly typical with $\mbf{y}_{l}$ is small for large $n$:
\begin{equation*}
\Pr\left[\mbf{Y}_{l}
\notin T^{(n)}_{\epsilon_{j+1}}\left(p \mid\Hat{\mbf{u}}^{\supset}_{\S_j},
\Hat{\mbf{u}}^{\ddag}_{\S_j,l} \mbf{u}_{\S_j}(\k_{\S_j}),\mbf{x}\right)\right]
\leq \delta_2 \left(n,\epsilon_j,\epsilon_{j+1},\mu(p)\right)\ .
\end{equation*}

An upper bound for the probability that there exists one or more codewords $\mbf{u}_{\S_j}(\tilde{\k}_{\S_j}) \neq \mbf{u}_{\S_j}(\k_{\S_j})$, which
satisfy \eqref{eq:decoder}, is
\begin{multline}\label{Sec:Proof:Eqn:Decoder-Error-Upperbound}
\Pr \left[\bigcup_{\set{K}_j}
\left\{
\Hat{\mbf{u}}_{\S_j}^{\supset},\Hat{\mbf{u}}_{\S_j,l}^{\ddag},
\mbf{y}_{l},\mbf{u}_{\S_j}(\tilde{\k}_{\S_j})
\right\}
\in T^{(n)}_{\epsilon_{j+1}}(p)\right]\\
< \exp_2 \Bigg[ n\Big(R'_{\S_j}+\sum_{i= |[l]\cap \S_j|+1}^{|\S_j|} R_{\S_j,i}
 - I\big(\aux_{\S_j} ;\set{A}^\supset_{\S_j}, \set{A}^\ddag_{\S_j,l}, Y_{l}\big)
 + 2\epsilon_{j+1}H\big(\aux_{\S_j}\big)\Big)\Bigg]\ ,
\end{multline}
where we have taken the union over
\begin{equation*}
\set{K}_j = \left\{\tilde{\k}_{\S_j} \neq
  \k_{\S_j},\ \{\tilde{k}_{\S_j,i} =
  k_{\S_j,i}\}_{i=1}^{|\{1,2,\ldots,l\}\cap \S_j|}\right\}\ .
\end{equation*}
Applying the union bound we get
\begin{align*}
\Pr \left[ D_{l,\S_j} \right]
&<\delta_2 +\exp_2\Bigg[n\Bigg(R_{\S_j}'+\sum_{i=|[l]\cap\S_j|}^{\S_j} R_{\S_j,i}\Bigg)
- n \Bigg( I\Big(U_{\S_j};\set{A}^\supset_{{\S_j}},\set{A}^\ddag_{{\S_j,l}},Y_l\Big) - 2 \epsilon_{j+1} H(U_{\S_j}) \Bigg) \Bigg]\ .
\end{align*}
Thus, if
\begin{equation}\label{Sec:7:Eqn:RL-SJ-Unique}
R'_{\S_j} + \sum_{i = |[l]\cap \S_j|+1}^{|\S_j|} R_{\S_j,i}
<I\left(\aux_{\S_j}; \set{A}^\supset_{\S_j}, \set{A}^\ddag_{\S_j,l},Y_{l}) - 2\epsilon_{j+1} H\left(\aux_{\S_j}\right)\right)
\end{equation}
then $\Pr[D_{l,\S_j}] \rightarrow 0$ as $n \rightarrow \infty$.

\subsection{Rate Constraints}

Consider decoder $l$ and any subset $\S_j$ where $l \in \S_j$. On combining the rate constraints~\eqref{Sec:7:Eqn:Error:E2-5}
and~\eqref{Sec:7:Eqn:RL-SJ-Unique} we get
\begin{align}
\notag
\sum_{i = 1}^{|\S \cap [l]|} R_{\S,i}
&> I\big(X,\set{A}^\supset_{\S_j},\set{A}^\dag_{\S_j};U_{\S}\big)
- I\big(U_{\S_j};\set{A}^\supset_{\S_j},\set{A}^\ddag_{\S_j,l},Y_l\big)\\
\label{Sec:Proof:Eqn:SJ-Unique-2}
&= I\big(X,\set{A}^\dag_{\S_j};U_{\S}\big|\set{A}^\supset_{\S_j}\big)
- I\big(U_{\S_j};\set{A}^\ddag_{\S_j,l},Y_l\big|\set{A}^\supset_{\S_j}\big)\ .
\end{align}
Since $\epsilon_j$ and $\epsilon_{j+1}$ may be selected arbitrarily small, we can ignore the $2(\epsilon_j + \epsilon_{j+1})H(\S_j)$ term.

Consider the other decoders in $[l] \cap \S_j$. Since $R_{\S_j,i} \geq 0$ for all $i$, it must be true that
\begin{align}
\notag
\sum_{i=1}^{|[l]\cap \S|} R_{\S_j,i}
&> \max_{\tilde{l} \in [l] \cap \S_j}
\Big[I\big(X,\set{A}^\dag_{\S_j};U_{\S_j}\big|\set{A}^\supset_{\S_j}\big)
- I\big(U_{\S_j};\set{A}^\ddag_{\S_j,\tilde{l}},Y_{\tilde{l}}\big|\set{A}^\supset_{\S_j}\big)\Big]\\
\label{Sec:7:Eqn:RL-SJ-Cor-Dec}
&= I\big(X,\set{A}^\dag_{\S_j};U_{\S_j}\big|\set{A}^\supset_{\S_j}\big)
- \min_{\tilde{l} \in [l] \cap \S_j} I\big(U_{\S_j};\set{A}^\ddag_{\S_j,\tilde{l}},Y_{\tilde{l}}\big|\set{A}^\supset_{\S_j}\big)\ ;
\end{align}
that is, the rate constraint for decoder $l$ must be at least as large as the rate constraint for decoder $\tilde{l}$ (for every $\tilde{l} \in [l]\cap\set{S}_j$).

The rate constraint~\eqref{Sec:7:Eqn:RL-SJ-Cor-Dec} is valid for any set $\S_j$ where $l \in \S_j$. For such subsets, define $l^* \triangleq \max_{i \in [l] \cap \S_j} i$. Since $l^* \in \S_j$ and $[l^*] \cap \S_j = [l] \cap \S_j$, it follows that~\eqref{Sec:7:Eqn:RL-SJ-Cor-Dec} is also valid for any set $\S_j$ where $[l] \cap \S_j \neq \emptyset$.

Finally, consider the sum rate $\sum_{i=1}^l R_i$ for the first $l$ channels. By construction, we have that
\begin{equation}\label{Sec:Proof:Eqn:Sum-Rate-1}
\sum_{i=1}^l R_i = \sum_{\S_j\subseteq [t] \atop{\S_j \cap [l] \neq\emptyset}} \sum_{i=1}^{|[l]\cap\S_j|} R_{\S_j,i}\ .
\end{equation}
Substituting the rate constraint~\eqref{Sec:7:Eqn:RL-SJ-Cor-Dec} into \eqref{Sec:Proof:Eqn:Sum-Rate-1} yields the desired result.

\section{$\epsilon$-Letter Typicality}\label{App:B}
For $\epsilon \geq 0$, a sequence $x^n \in \set{X}^n$ is said to be $\epsilon$-letter typical with respect to a discrete memoryless source $(\set{X},p_X)$ if
\begin{equation*}
\left|\frac{1}{n}N(a|x^n) - p_X(a) \right| \leq \epsilon \cdot
p_X(a)\quad \forall a \in \set{X}\ ,
\end{equation*}
where $N(a|x^n)$ is the number of times the letter $a$ occurs in the
sequence $x^n$. The collection of all $\epsilon$-letter typical
sequences is denoted by $T^{(n)}_{\epsilon}(p_X)$.

In a similar fashion, a pair of sequences $x^n$ and $y^n$ are said to
jointly $\epsilon$-letter typical with respect to a discrete
memoryless two source $(\set{X} \times \set{Y},p_{XY})$ if
\begin{equation*}
\left|\frac{1}{n}N(a,b|x^n,y^n) - p_{XY}(a,b) \right| \leq \epsilon \cdot p_{XY}(a,b)\quad \forall (a,b) \in \set{X} \times \set{Y}\ ,
\end{equation*}
where $N(a,b|x^n,y^n)$ is the number of times the pair of letters
$(a,b)$ occurs in the pair $(x^n,y^n)$. The collection of all joint
$\epsilon$-typical sequence pairs is denoted by
$T_\epsilon^{(n)}(p_{XY})$.

Given $(\set{X} \times \set{Y},p_{XY})$ and $x^n \in \set{X}^n$, the set
\begin{equation*}
T_\epsilon^{(n)}\left(p_{XY}\mid x^n\right) = \big\{ y^n\ :\ (x^n,y^n) \in T_\epsilon^{(n)}(p_{XY}) \big\}
\end{equation*}
is called the set of conditionally $\epsilon$-letter typical sequences.

Let
$\mu(\set{X},p_{X}) = \min \{ p_{X}(x) :\ x \in \text{support}(p_{X})\}$
and define
\begin{equation*}
\delta_1\left(n,\epsilon,\mu(p_X)\right) = 2|\set{X}| \cdot
e^{-n\epsilon^2\mu(p_{X})}.
\end{equation*}
Note, $\delta\big(n,\epsilon,\mu(p_X)\big) \rightarrow 0$ as $n \rightarrow \infty$.

\begin{lemma}[Theorem 1.1, \cite{Kramer-2008-A}]\label{App:Lem:1}
Suppose $X^n$ is emitted by a discrete memoryless source $(\set{X},p_X)$. If $0 < \epsilon \leq \mu(p_{X})$, then
\begin{equation*}
1 - \delta_1\left(n,\epsilon,\mu(p_X)\right) \leq \Pr\left[X^n \in T^{(n)}_{\epsilon}(p_{X}) \right] \leq 1\ .
\end{equation*}
\end{lemma}

Now consider a discrete memoryless two-source $(\set{X} \times
\set{Y},p_{XY})$, let
\begin{equation*}
\delta_2\big(n,\epsilon_1,\epsilon_2,\mu(p_{XY})\big) =
2|\set{X}||\set{Y}| \cdot e^{-n\frac{(\epsilon_2 -
    \epsilon_1)^2}{1+\epsilon_1}\mu(p_{XY})},
\end{equation*}
and note that $\delta_2\big(n,\epsilon_1,\epsilon_2,\mu(p_X)\big)
\rightarrow 0$ as $n \rightarrow \infty$.
\begin{lemma}[Theorem 1.3, \cite{Kramer-2008-A}]\label{App:Lem:2}
Suppose $Y^n$ is emitted by $(\set{Y},p_Y)$ where $p_Y$ is equal to the $Y$-marginal of $p_{XY}$. If $0 < \epsilon_1 < \epsilon_2 \leq \mu(p_{XY})$ and $x^n \in T^{(n)}_{\epsilon_1}(p_{X})$, then
\begin{multline*}
\left(1 - \delta_2\left(n,\epsilon_1,\epsilon_2,\mu(p_{XY})\right)\right) 2^{-n\left(I(X;Y) + 2\epsilon_2H(Y)\right)} \\
\leq \Pr\left[ Y^n \in T^{(n)}_{\epsilon_2}\left(p_{XY}\mid x^n\right) \right] \leq 2^{-n\left(I(X;Y)-2 \epsilon_2 H(Y) \right)} .
\end{multline*}
\end{lemma}

Finally, a direct consequence of Lemma~\ref{App:Lem:2} for Markov
sources is the following result.
\begin{lemma}[Markov Lemma~\cite{Kramer-2008-A}]\label{App:Lem:3}
  Suppose $(X^n,Y^n,Z^n)$ is emitted by a discrete memoryless
  three-source $(\set{X} \times \set{Y} \times \set{Z},p_{XYZ})$ where
  $X \markov Y \markov Z$. If $0 < \epsilon_1 < \epsilon_2 \leq
  \mu(p_{XYZ})$ and $(x^n,y^n) \in T^{(n)}_{\epsilon_1}(p_{XY})$, then
\begin{align*}
\Pr&\left[Z^n \in T_{\epsilon_2}^{(n)} \left( p_{XYZ} \mid x^n,y^n
  \right) \mid Y^n = y^n\right]\\
&= \Pr\left[Z^n \in T_{\epsilon_2}^{(n)}\left(p_{XYZ} \mid
    x^n,y^n\right) \mid X^n = x^n,Y^n = y^n\right]\\
&\geq 1-\delta_2\left(n,\epsilon_1,\epsilon_2,\mu(p_{XYZ})\right).
\end{align*}
\end{lemma}
\bibliographystyle{IEEEtran}

\begin{thebibliography}{10}
\providecommand{\url}[1]{#1}
\csname url@samestyle\endcsname
\providecommand{\newblock}{\relax}
\providecommand{\bibinfo}[2]{#2}
\providecommand{\BIBentrySTDinterwordspacing}{\spaceskip=0pt\relax}
\providecommand{\BIBentryALTinterwordstretchfactor}{4}
\providecommand{\BIBentryALTinterwordspacing}{\spaceskip=\fontdimen2\font plus
\BIBentryALTinterwordstretchfactor\fontdimen3\font minus
  \fontdimen4\font\relax}
\providecommand{\BIBforeignlanguage}[2]{{%
\expandafter\ifx\csname l@#1\endcsname\relax
\typeout{** WARNING: IEEEtran.bst: No hyphenation pattern has been}%
\typeout{** loaded for the language `#1'. Using the pattern for}%
\typeout{** the default language instead.}%
\else
\language=\csname l@#1\endcsname
\fi
#2}}
\providecommand{\BIBdecl}{\relax}
\BIBdecl

\bibitem{Wyner-Jan-1976-A}
A.~Wyner and J.~Ziv, ``The {R}ate-{D}istortion {F}unction for {S}ource {C}oding
  with {S}ide {I}nformation at the {D}ecoder,'' \emph{IEEE Transactions on
  Information Theory}, vol.~22, no.~1, pp. 1--10, 1976.

\bibitem{Csiszar-1981-B}
I.~Csisz$\acute{\text{a}}$r and J.~K$\ddot{\text{o}}$rner, \emph{Information
  {T}heory: {C}oding {T}heorems for {D}iscrete {M}emoryless {S}ystems}.\hskip
  1em plus 0.5em minus 0.4em\relax Academic Press, 1981.

\bibitem{Cover-1991-B}
T.~Cover and J.~Thomas, \emph{Elements of {I}nformation {T}heory}.\hskip 1em
  plus 0.5em minus 0.4em\relax New York: Wiley, 1991.

\bibitem{Kaspi-Nov-1994-A}
A.~H. Kaspi, ``Rate-{D}istortion {F}unction when {S}ide-{I}nformation {M}ay
  {B}e {P}resent at the {D}ecoder,'' \emph{IEEE Transactions on Information
  Theory}, vol.~40, no.~6, pp. 2031--2034, 1994.

\bibitem{Heegard-Nov-1985-A}
C.~Heegard and T.~Berger, ``Rate {D}istortion when {S}ide {I}nformation {M}ay
  {B}e {A}bsent,'' \emph{IEEE Transactions on Information Theory}, vol.~31,
  no.~6, pp. 727--734, 1985.

\bibitem{Tian-Aug-2007-A}
C.~Tian and S.~Diggavi, ``On {M}ultistage {S}uccessive {R}efinement for
  {W}yner-{Z}iv {S}ource {C}oding with {D}egraded {S}ide {I}nformations,''
  \emph{IEEE Transactions on Information Theory}, vol.~53, no.~8, pp.
  2946--2960, 2007.

\bibitem{Tian-Dec-2008-A}
C.~Tian and S.~N. Diggavi, ``Side-{I}nformation {S}calable {S}ource {C}oding,''
  \emph{IEEE Transactions on Information Theory}, vol.~54, no.~12, pp.
  5591--5608, 2008.

\bibitem{Sgarro-Mar-1977-A}
A.~Sgarro, ``Source {C}oding with {S}ide {I}nformation at {S}everal
  {D}ecoders,'' \emph{IEEE Transactions on Information Theory}, vol.~23, no.~2,
  pp. 179--182, 1977.

\bibitem{Steinberg-Aug-2004-A}
Y.~Steinberg and N.~Merhav, ``On {S}uccessive {R}efinement for the
  {W}yner-{Z}iv {P}roblem,'' \emph{IEEE Transactions on Information Theory},
  vol.~50, no.~8, pp. 1636--1654, 2004.

\bibitem{Tian-Oct-2006-C}
C.~Tian and S.~N. Diggavi, ``A {C}alculation of the {H}eegard-{B}erger
  {R}ate-{D}istortion {F}unction for a {B}inary {S}ource,'' in
  \emph{proceedings of the IEEE Information Theory Workshop}, Chengdu, China,
  2006, pp. 342--346.

\bibitem{Yeung-2002-B}
R.~Yeung, \emph{A {F}irst {C}ourse in {I}nformation {T}heory}.\hskip 1em plus
  0.5em minus 0.4em\relax Kluwer Academic/Plenum Publishers, 2002.

\bibitem{Gray-Nov-1974-A}
R.~Gray and A.~Wyner, ``Source {C}oding for a {S}imple network,'' \emph{Bell
  System Technical Journal}, vol.~53, no.~9, pp. 1681--1721, 1974.

\bibitem{Effros-Sep-1999-A}
M.~Effros, ``Distortion-{R}ate {B}ounds for {F}ixed-and {V}ariable-{R}ate
  {M}ultiresolution {S}ource {C}odes,'' \emph{IEEE Transactions on Information
  Theory}, vol.~45, no.~6, pp. 1887--1910, 1999.

\bibitem{Tian-Jul-2009-C}
C.~Tian, ``Latent {C}apacity {R}egion: {A} {C}ase {S}tudy on {S}ymmetric
  {B}roadcast with {C}ommon {M}essages,'' in \emph{proceedings of the IEEE
  International Symposium on Information Theory}, Seoul, Korea, 2009, pp.
  1834--1838.

\bibitem{Vellambi-Mar-2010-C}
B.~N. Vellambi and R.~Timo, ``Multi-{T}erminal {S}ource {C}oding: {C}an
  {Z}ero-{R}ate {E}ncoders {E}nlarge the {R}ate {R}egion?'' in
  \emph{proceedings of the International Zurich Seminar on Communications},
  Zurich, Switzerland, 2010.

\bibitem{Rockafellar-1997-B}
R.~T. Rockafellar, \emph{Convex {A}nalysis}.\hskip 1em plus 0.5em minus
  0.4em\relax Princeton {U}niversity {P}ress, 1997.

\bibitem{Slepian-Jul-1973-A}
D.~Slepian and J.~Wolf, ``Noiseless {C}oding of {C}orrelated {I}nformation
  {S}ources,'' \emph{IEEE Transactions on information Theory}, vol.~19, no.~4,
  pp. 471--480, 1973.

\bibitem{Bakshi-Jul-2008-C}
M.~Bakshi and M.~Effros, ``On {A}chievable {R}ates for {M}ulticast in the
  {P}resence of {S}ide {I}nformation,'' in \emph{proceeding of the IEEE
  International Symposium on Information Theory}, Toronto, Canada, 2008, p.
  1661–1665.

\bibitem{Fu-July-2002-A}
F.~Fang-Wei and R.~W. Yeung, ``On the {R}ate-{D}istortion {R}egion for
  {M}ultiple {D}escriptions,'' \emph{IEEE Transactions on Information Theory},
  vol.~48, no.~7, pp. 2012--2021, 2002.

\bibitem{Kramer-2008-A}
G.~Kramer, ``Topics in {M}ulti-{U}ser {I}nformation {T}heory,''
  \emph{Foundations and Trends in Communications and Information Theory},
  vol.~4, no. 4–5, pp. 265--444, 2008.

\end{thebibliography}

\end{document}